\documentclass[final,5p,times,twocolumn]{elsarticle}
\newif\ifjournal
\journalfalse

\ifjournal\else
\fi

\usepackage[utf8]{inputenc}
\usepackage[T1]{fontenc}
\usepackage{lmodern} % scalable Type 1 text fonts (no bitmap Type 3 fonts in the PDF)
\usepackage{amsmath,amssymb}
\usepackage{algorithm}
\usepackage[noend]{algpseudocode}
\usepackage{graphicx}
\usepackage{booktabs}
\usepackage{array}
\usepackage{rotating}
\usepackage{adjustbox}
\usepackage{float}
\usepackage{pdflscape}
\usepackage{longtable}
\usepackage{placeins} % floats stay within their section
\usepackage{multirow}
\usepackage{xcolor}
\usepackage{lineno}
\ifjournal\usepackage{xr-hyper}\fi
\usepackage[colorlinks=true,linkcolor=blue!60!black,citecolor=blue!60!black,urlcolor=blue!60!black]{hyperref}

\newcommand{\gscn}{\mbox{CGS}}
\newlength{\colfigw}\AtBeginDocument{\ifdim\columnwidth<0.7\textwidth\setlength{\colfigw}{\columnwidth}\else\setlength{\colfigw}{0.6\textwidth}\fi}
\newsavebox{\fitbox}
\newcommand{\curtable}{?}
\newcommand{\fitheight}{0.8\textheight}  % room a table may take below its caption (0.8\textwidth on a landscape page)
\newcommand{\fitwidth}[1]{\sbox{\fitbox}{#1}\typeout{TABLEWIDTH \curtable\space\the\wd\fitbox\space\the\linewidth\space\the\dimexpr\ht\fitbox+\dp\fitbox\relax\space\the\textwidth\space\the\textheight}%
  \adjustbox{max width=\linewidth,max totalheight=\fitheight}{\usebox{\fitbox}}}
\newcommand{\rotatedtables}{}
\newif\ifrotatedtable
\newif\ifptablehere  % journal appendices: each table right below its heading
\newcommand{\ptablehook}{}  % arXiv: an appendix heading set at the head of its first table
\makeatletter
\newenvironment{ptable}[2][tbp]{\renewcommand{\curtable}{#2}%
  \edef\@tempa{\noexpand\in@{,#2,}{,\rotatedtables,}}\@tempa
  \ifptablehere\rotatedtablefalse\begin{table}[H]%
  \else\ifin@\rotatedtabletrue\begin{sidewaystable}\renewcommand{\fitheight}{0.8\textwidth}%
  \else\rotatedtablefalse\begin{table*}[#1]\fi\fi
  \def\baselinestretch{1}\normalsize\ptablehook}% tables single-spaced, also in the review layout
 {\ifptablehere\end{table}\else\ifrotatedtable\end{sidewaystable}\else\end{table*}\fi\fi}
\makeatother
\makeatletter\newcommand{\sref}[1]{\@ifundefined{supp@#1}{\ref{#1}}{\@nameuse{supp@#1}}}\makeatother\input{tables/supp_labels_arxiv}  % material that moves to an appendix in the journal build
\makeatletter\newcommand{\ssecref}[1]{\@ifundefined{supp@#1}{\S\ref{#1}}{Supplementary \S\@nameuse{supp@#1}}}\makeatother  % a section that becomes an appendix (its number reads "Appendix X")
\newcommand{\best}[1]{\textbf{#1}}

\newcommand{\pend}{\mbox{\textcolor{red}{pending}}}

\makeatletter
\newcommand{\num}[1]{\@ifundefined{num@#1}{\pend}{\@nameuse{num@#1}}}
\makeatother
\newcommand{\numm}[1]{\ensuremath{\num{#1}}}
\input{tables/numbers}
\input{tables/budget_numbers}
\input{tables/v4_numbers}
\input{tables/full_numbers}
\input{tables/rollout_numbers}
\input{tables/extra_numbers}
\makeatletter
\@namedef{num@lol.units}{57}
\@namedef{num@lol.wins}{57}
\@namedef{num@lol.median}{4.9\times10^{3}}
\@namedef{num@lol.min}{3.4}
\@namedef{num@lol.max}{1.4\times10^{6}}
\@namedef{num@lol.min.unit}{series~3}
\@namedef{num@lol.runs}{114}
\@namedef{num@lol.capped}{24}
\@namedef{num@lol.qbetter}{38}
\makeatother

\input{tables/traj_numbers}
\input{tables/kink_numbers}
\input{tables/noise_transition_numbers}
\input{tables/system_numbers}

\journal{Chaos, Solitons and Fractals}
\ifjournal\else\makeatletter\def\ps@pprintTitle{\let\@oddhead\@empty\let\@evenhead\@empty\def\@oddfoot{\footnotesize\itshape Preprint\hfill\today}\let\@evenfoot\@oddfoot}\makeatother\fi
\renewcommand{\dbltopfraction}{0.85}\renewcommand{\dblfloatpagefraction}{0.8}\predisplaypenalty=0\makeatletter\g@addto@macro\normalsize{\setlength\abovedisplayskip{5pt plus 2pt minus 2pt}\setlength\belowdisplayskip{5pt plus 2pt minus 2pt}\setlength\abovedisplayshortskip{0pt plus 2pt}\setlength\belowdisplayshortskip{3pt plus 2pt minus 1pt}}\makeatother\makeatletter\renewcommand\section{\@startsection{section}{1}{\z@}{10\p@ \@plus 4\p@ \@minus 2\p@}{5\p@ \@plus 2\p@ \@minus 1\p@}{\normalsize\bfseries\boldmath}}\renewcommand\subsection{\@startsection{subsection}{2}{\z@}{7\p@ \@plus 3\p@ \@minus 2\p@}{2\p@ \@plus 2\p@ \@minus 1\p@}{\normalfont\normalsize\itshape}}\renewcommand\elsparagraph{\@startsection{paragraph}{4}{0\z@}{5\p@ \@plus 3\p@ \@minus 2\p@}{-6\p@}{\normalfont\itshape}}\makeatother
\begin{document}

\begin{frontmatter}

\title{Chaos Gaussian Splatting: Polynomial Regression with Gaussian Splats for Chaotic Maps with Gradient Discontinuities}

\author[usp]{Daniel Ortiz Valete\corref{cor1}}
\ead{valeda02@usp.br}
\author[usp]{Gonzalo Travieso}
\ead{gonzalo@ifsc.usp.br}
\author[usp]{Odemir Martinez Bruno}
\ead{bruno@ifsc.usp.br}
\affiliation[usp]{organization={S\~{a}o Carlos Institute of Physics, University of S\~{a}o Paulo},
            addressline={Avenida Trabalhador S\~{a}o-carlense 400},
            city={S\~{a}o Carlos},
            postcode={13566-590},
            state={SP},
            country={Brazil}}

\begin{abstract}
Polynomial regression forecasts noise-free, fully observed, low-dimensional chaotic systems almost to machine
precision, but many one-step maps are not low-degree polynomials: maps with kinks, and flows observed through a single
coordinate. We introduce chaos Gaussian splatting (\gscn{}), which models the map as a
polynomial trend plus Gaussian ``splats'', local quadratics split and pruned as in Gaussian splatting and fitted in
closed form, reducing to polynomial regression where that suffices. Compact mode fits one splat field; full mode, for
kinked maps, gives the splats hinges, combines six splat fields and defers, where they disagree, to a referee that
locates kinks in the training data. In a pre-registered test on fresh initial conditions of ten systems
(\num{f910.units} units), full mode had a lower one-step error than the S-map and the Farmer--Sidorowich predictor, each
given every input view \gscn{} uses, on every unit, with a median error \num{f910.ratio.medianinv} times lower; it
predicts \num{f910.pred.ratio} times more slowly. In closed loop (post hoc) its forecasts outlasted those methods' and an echo
state network's on every map unit, by \num{roll.full.lead.min} to \num{roll.full.lead.max} steps. On six scalar kinked maps with $10^5$ points, compact mode met an accuracy target set
by the Farmer--Sidorowich predictor in \num{bud.k1.100000.val} of \num{bud.k1.100000.runs} runs and then used
\num{bud.k1.100000.gm.min}--\num{bud.k1.100000.gm.max} times less memory than any local model meeting it. Once
observation noise dominates, the methods' errors agree within a factor of 1.5 (\num{nt.h1.within} of
\num{nt.h1.cells} cases, a second pre-registered study); on a measured laser series \gscn{} was not more accurate than
the S-map.
\end{abstract}


\end{frontmatter}
\hypersetup{pdfauthor={Daniel Ortiz Valete, Gonzalo Travieso, Odemir Martinez Bruno}}

\ifjournal\linenumbers\fi

%% ====================================================================
\section{Introduction}

\begin{figure*}[t]
\centering
\includegraphics[width=\textwidth]{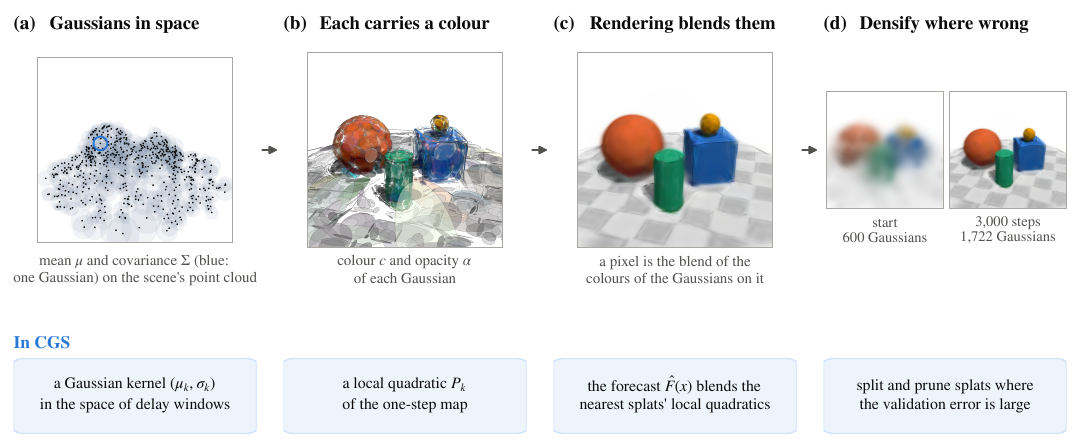}
\caption{Three-dimensional Gaussian splatting~\cite{kerbl20233d}, on one small run on a synthetic scene (ray-traced
photographs from 20 cameras), and the corresponding element of \gscn{} below each step. (a)~A scene is a set of 3D
Gaussians, each with a mean and a covariance, initialised on the point cloud recovered from the photographs; in \gscn{},
Gaussian kernels in the space of delay windows. (b)~Each Gaussian carries a colour and an opacity; each splat of
\gscn{} carries a local quadratic of the one-step map. (c)~A pixel's colour is the front-to-back alpha blend of the
Gaussians covering it; a forecast of \gscn{} is the normalised blend of the nearest splats' local quadratics.
(d)~The Gaussians are optimised so that the renders match the photographs, and adaptive density control splits and
prunes them (the render of the initial 600 Gaussians and after 3{,}000 steps, 1{,}722 Gaussians); \gscn{} splits and
prunes its splats where the validation error is large, and fits them in closed form rather than by gradient descent.}
\label{fig:3dgs}
\end{figure*}

\paragraph{Why chaotic forecasting needs precise maps} In chaotic systems small forecast errors grow at the rate of
the largest Lyapunov exponent $\lambda$~\cite{ott2002chaos,lorenz1969predictability}, so a recursive forecast stays
useful for a number of steps that grows only with the logarithm of the one-step error: every factor gained in the
learned one-step map buys forecast time. Many systems of interest are piecewise smooth: switching, impacting and
saturating systems have maps that are continuous but whose derivative jumps on a boundary, a \emph{kink}~\cite{dibernardo2008piecewise}.

\paragraph{Limits of global and local models} On noise-free, fully observed low-dimensional systems a global polynomial
is highly accurate: PolyProp~\cite{polyprop2025}, fitted in 512-bit arithmetic, matches a 64-bit solver of the true
equations. Other data-driven forecasters include recurrent and attention-based
networks~\cite{hochreiter1997long,lai2018modeling,vlachas2018data,vlachas2020backpropagation,valle2025forecasting},
reservoir computers (ESN, NG-RC)~\cite{jaeger2004harnessing,pathak2017using,pathak2018model,gauthier2021next}, sparse
regression of governing equations~\cite{brunton2016discovering} and pre-trained foundation
models~\cite{ansari2024chronos,zhang2025zeroshot}. Global models fail on maps with kinks, where the best of them stays
at least $\numm{kink.globalfloor.min}$ times above the float32 rounding error (\S\ref{sec:maps}), and on flows observed
through one coordinate, whose delay map is smooth~\cite{takens1981detecting} but in general not a
polynomial~\cite{letellier1998nonequivalence} (\ssecref{sec:flows}). Local approximation in reconstructed state
space~\cite{farmer1987predicting,casdagli1989nonlinear,sugihara1994nonlinear} handles kinks, but it keeps every training
point and fits a model at every query, and near a kink its error falls only slowly with the data (\S\ref{sec:discussion}).
Two references are kept apart from what a model reaches: the \emph{rounding floor}, the mean squared difference between
the exact next value and its float32 rounding (the error of a perfect forecast scored against float32 targets), as in
Valle and Bruno's benchmark~\cite{valle2025forecasting}, and the \emph{exact map} applied to the float32 inputs. Neither
is a lower bound: a model reading a window of past values can recover part of its inputs' rounding, as full mode does
on most maps (\S\ref{sec:maps}).

\begin{figure*}[tbp]
  \centering\ifjournal\def\baselinestretch{1}\normalsize\fi % single-spaced caption: room for the figure
  \includegraphics[width=\textwidth]{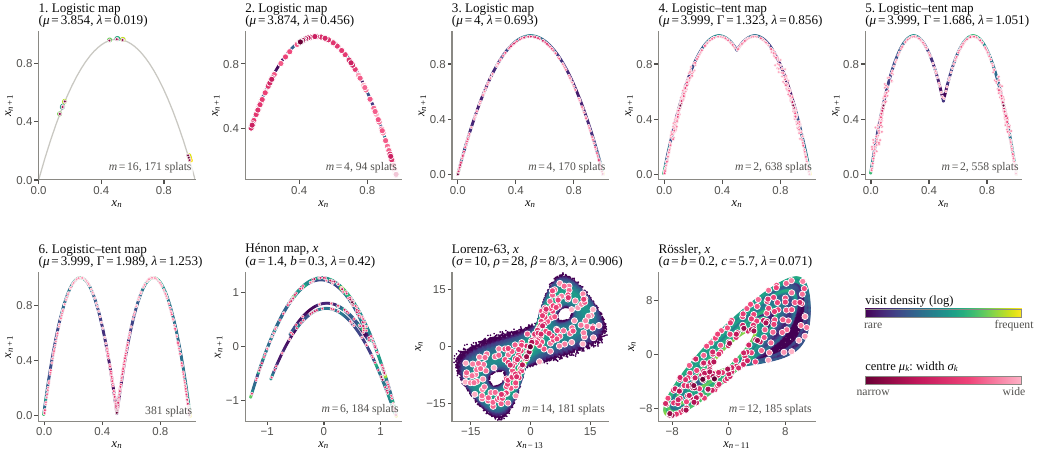}
  \caption{The attractor of each benchmark series with the Gaussian centres $\mu_k$ compact \gscn{} fitted to
  it; panels 1--6 are Valle and Bruno's series~\cite{valle2025forecasting}, each with its map's parameters and largest
  Lyapunov exponent $\lambda$ (per step; per time unit for the flows). Background: observed points coloured by visit density (log
  scale, violet rare, yellow frequent). Markers: splat centres, darker = narrower $\sigma_k$; lower right, the validated
  window $m$ and the number of splats. Maps and H\'enon: the plane of consecutive observations, centres projected from
  the window (for $m=1$ placed on the learned map, $(\mu_k, F(\mu_k))$); flows: against the value $m-1$ steps earlier.
  Series~1 is chaotic on \num{s1b.bands} narrow bands (together \num{s1b.width} wide) visited in a fixed cycle, dots on the
  logistic map's graph (grey); all \num{s1b.inband} centres lie on the bands, as do the forecasts (\ssecref{sec:longr2}).}
  \label{fig:attractors}
\end{figure*}

\paragraph{Adaptive splatting} 3D Gaussian splatting~\cite{kerbl20233d} represents a scene by Gaussians that it splits
and prunes where the render is wrong; splatting itself renders a volume as overlapping
footprints~\cite{westover1990footprint}, and TimeGS~\cite{wang2026forecasting} brought it to time-series forecasting as
rendering on a period--phase grid. \gscn{} brings the same idea to the one-step map of a chaotic system
(Fig.~\ref{fig:3dgs}): each Gaussian splat in delay space carries a local quadratic of the map over a polynomial trend,
all fits are closed-form least squares, splats are split and pruned where the residual requires, and every choice is
made on validation data (Fig.~\ref{fig:attractors}). Its parts have precedents: radial-basis-function networks place
Gaussian units in delay space~\cite{broomhead1988multivariable,casdagli1989nonlinear,platt1991resource}; normalised
Gaussian units carry local linear
models~\cite{stokbro1990exploiting,takagi1985fuzzy,nelles1996basis,vijayakumar2005incremental}; splat
regression~\cite{daniels2026splat} fits splats by a gradient flow; and kinks have been modelled by hinges, splines,
piecewise-affine identification and implicit sparse
regression~\cite{breiman1993hinging,friedman1991multivariate,paoletti2007identification,kaheman2020sindy,li2026image}.
Their combination is new: Gaussians split and pruned by error density and opacity as in graphics, each holding a hinged
local model that can represent a kink exactly, over a polynomial trend to which the model reduces where that suffices.

\begin{figure*}[tbp]
\centering\includegraphics[width=0.82\textwidth]{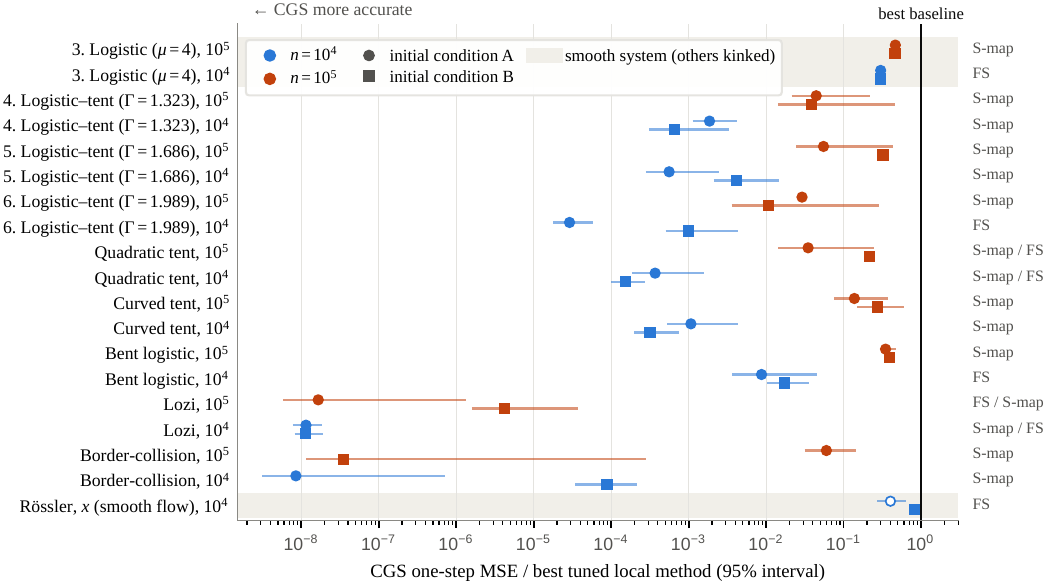}
\caption{The main result: the pre-registered confirmation of full mode (\S\ref{sec:fresh}). Each row is one
\emph{unit}, a system at one training size ($n=10^4$, blue; $10^5$, orange), with one point per fresh initial condition
(A, circles; B, squares). A point is \gscn{}'s one-step test MSE divided by that of the better of the two tuned local
methods, the S-map and the Farmer--Sidorowich predictor (FS), each given every input view; the right column names that
baseline. The horizontal bars are 95\% moving-block bootstrap intervals of the ratio (blocks of 500 test points,
2\,000 resamples); a unit is a resolved win when its whole interval lies left of the vertical line at 1. The axis is
logarithmic: a point at $10^{-6}$ is a million times more accurate than the better baseline. Shaded rows are the smooth
systems (series 3 and R\"ossler), where the margins are smallest; the other maps are kinked. Hollow: baseline job
completed after Kaggle's time limit. Series as in Table~\ref{tab:maps}; values in Table~\sref{tab:fresh910}.}
\label{fig:fresh}
\end{figure*}

\paragraph{Contributions} (i)~\emph{Method:} an adaptive splat regression, in a compact mode (one splat field) and a
full mode (hinged fields combined, with a network referee where they disagree), fitted without gradient descent
(\S\ref{sec:method}). (ii)~\emph{Representation:} on kinked maps with $10^5$ or more points compact mode matches the
local methods' accuracy with $\num{bud.k1.100000.gm.min}$--$\num{bud.k1.900000.gm.max}$ times less memory, because its
size does not grow with the data (\S\S\ref{sec:cost}--\ref{sec:budget}). (iii)~\emph{Dynamics:} kinks are treated
explicitly, by hinges and by locating them in the training windows' neighbour graph; we show where the error of local
methods sits near a kink and how the one-step precision gained translates into forecast horizon through the Lyapunov
exponent (\S\ref{sec:discussion}). We test \gscn{} on Valle and Bruno's six map series~\cite{valle2025forecasting}, the
H\'enon map~\cite{henon1976two}, Lorenz-63~\cite{lorenz1963deterministic} and R\"ossler~\cite{rossler1976equation}
observed through $x$, the measured Santa Fe laser~\cite{weigend1994time} and six further kinked maps in two
pre-registered tests (the dysts benchmark~\cite{gilpin2021chaos} and full-state Lorenz-63 and
Lorenz-96~\cite{lorenz1996predictability} are in the supplement), and confirm full mode in a pre-registered test on fresh
initial conditions, where it beat the S-map and the Farmer--Sidorowich predictor on \num{f910.unitwins} of
\num{f910.units} units (\S\ref{sec:fresh}, Fig.~\ref{fig:fresh}).

%% ====================================================================
\section{The CGS model}\label{sec:method}

\gscn{} runs in two modes: \emph{compact}, one splat field whose size and prediction cost do not grow with the data
(\S\S\ref{sec:field}--\ref{sec:selection}; its cost in \S\ref{sec:cost}), and \emph{full}, the mode confirmed on fresh data (\S\ref{sec:fresh}), several
splat fields combined by their medoid or, where they disagree, by a network referee (\S\ref{sec:v4}). The fields are, on a
kinked map, hinged fields at three seeds, with and without colour channels; on a smooth map, the median of three seeds'
fits; on a finely sampled flow, the mean over several views of the history. Full-state series use compact mode with
closed-loop selection (\ssecref{sec:closedloop}).

\subsection{Problem setting and notation}\label{sec:setting}
A scalar series gives delay vectors
\begin{equation}
  x_t = (s_t, s_{t-1}, \dots, s_{t-m+1})^{\top} \in \mathbb{R}^m
  \label{eq:delay}
\end{equation}
and targets $y_t = s_{t+1}$; \gscn{} learns $F$ with $F(x_t)\approx y_t$ on training pairs
$\mathcal{D}=\{(x_i,y_i)\}_{i=1}^{N}$, makes every choice on later validation pairs $\mathcal{V}$, and is scored on a
test segment. Forecasts are \emph{recursive}: each prediction, rounded to float32, is fed back. For an observed
full state $z_t\in\mathbb{R}^d$, $x_t=z_t$ (or a window of the last $k$ states, \ssecref{sec:closedloop}) and
$F:\mathbb{R}^{d}\to\mathbb{R}^{d}$, every formula below applying componentwise.

\subsection{Compact mode: the splat field}\label{sec:field}
Compact mode models $F$ as a polynomial trend plus a \emph{splat field} of $K$ Gaussian splats,
\begin{equation}
  F(x) = T(x) + \sum_{k\in\mathcal{N}(x)} a_k(x)\,P_k\big(u_k(x)\big),
  \label{eq:field}
\end{equation}
over the $n_b=4$ centres $\mu_k$ nearest to $x$, with widths $\sigma_k$, local coordinates $u_k(x)=(x-\mu_k)/\sigma_k$
and blending weights
\begin{equation}
  a_k(x) = \frac{\exp\!\big(-\tfrac12\|u_k(x)\|^2\big)}{\sum_{j\in\mathcal{N}(x)}\exp\!\big(-\tfrac12\|u_j(x)\|^2\big)},
  \qquad k\in\mathcal{N}(x),
  \label{eq:weights}
\end{equation}
a partition of unity. Each splat carries a quadratic in its local coordinates,
\begin{equation}
  P_k(u) = c_{k,0} + \sum_{i=1}^{m} c_{k,i}\,u_i + \sum_{1\le i\le j\le m} c_{k,ij}\,u_i u_j ,
  \label{eq:local}
\end{equation}
with $p=1+m+m(m+1)/2$ coefficients; $\sigma_k$ is the mean distance from $\mu_k$ to its two nearest centres. With
$\phi_k(x)$ the monomials of Eq.~\eqref{eq:local} at $u_k(x)$,
\begin{equation}
  F(x) = T(x) + \sum_{k\in\mathcal{N}(x)} a_k(x)\,\phi_k(x)^{\top}c_k = T(x) + \Phi(x)^{\top}c ,
  \label{eq:linear}
\end{equation}
linear in all coefficients for fixed centres, so fitting is sparse least squares. Unlike the primitives of
Gaussian splatting in graphics~\cite{kerbl20233d} and time-series forecasting~\cite{wang2026forecasting}, each splat
carries a local model of the map. $\Phi(x)\in\mathbb{R}^{Kp}$ is non-zero only in the four neighbours' blocks
$a_k(x)\phi_k(x)$. With $c=0$, Eq.~\eqref{eq:field} is the global polynomial $T$, the model class of polynomial
regression, NG-RC and PolyProp (\ssecref{sec:polyrelation}); with $T\equiv0$, a pure splat field. Both are candidates
decided on validation: on the smooth full-state flows validation keeps the trend alone, and on the maps that are
low-degree polynomials (logistic, H\'enon) the trend alone is within 20 per cent of the selected model
(Table~\sref{tab:regimes}).

\subsection{Global polynomial trend}\label{sec:trend}
The trend is a polynomial of total degree $q$ in standardised inputs $\tilde x$,
\begin{equation}
  T(x) = \sum_{|\alpha|\le q} \beta_\alpha\, \tilde x^{\alpha} = \psi_q(x)^{\top}\beta,
  \label{eq:trend}
\end{equation}
fitted by ridge regression scaled to the Gram matrix $A=\sum_i \psi_q(x_i)\psi_q(x_i)^{\top}$,
\begin{equation}
  \hat\beta(q,\lambda) = \Big(A + \lambda\,\tfrac{\operatorname{tr}A}{D_q}\,I\Big)^{-1}\sum_i \psi_q(x_i)\,y_i ,
  \label{eq:trend-ridge}
\end{equation}
with $(q,\lambda)$ chosen on validation. The trend has $D_q=\tbinom{m+q}{q}$ monomials; $\lambda\in\{0, 10^{-14},
10^{-12},\dots,10^{-2}, 1\}$; degrees $q=1,\dots,8$ while $D_q\le N/2$, $D_q\le 1.2\times10^4$ and
$N D_q\le1.5\times10^{9}$. The same regression, on a window it selects itself, is the polynomial baseline
(\S\ref{sec:baselines}).

\subsection{Coefficient estimation}\label{sec:solvers}
The splats are fitted to the trend residuals $r_i = y_i - T(x_i)$, either jointly (the \emph{global solver}),
\begin{equation}
  \begin{aligned}
  \hat c &= \arg\min_{c}\ \sum_{i=1}^{N}\big(\Phi(x_i)^{\top}c - r_i\big)^2 + \lambda_g\|c\|^2\\
         &= \big(G+\lambda_g I\big)^{-1} b,
  \end{aligned}
  \label{eq:global}
\end{equation}
or splat by splat on the points $I_k=\{i:\,k\in\mathcal{N}(x_i)\}$ each touches, weighted by its blending weights (the \emph{local solver}),
\begin{equation}
  \begin{aligned}
  \hat c_k &= \arg\min_{c}\ \sum_{i\in I_k} a_k(x_i)\big(\phi_k(x_i)^{\top}c - r_i\big)^2\\
  &\qquad + \lambda_\ell\,\tfrac{\operatorname{tr}G_k}{p}\,\|c\|^2 .
  \end{aligned}
  \label{eq:local-fit}
\end{equation}
The local solver is always tried, the global one only if the coupling degree $\kappa$ (the mean number of splats sharing
a training point with a given splat) is at most 25. Here $G=\sum_i\Phi(x_i)\Phi(x_i)^{\top}$,
$b=\sum_i\Phi(x_i)\,r_i$, $\lambda_g=10^{-12}$; $G$ is block-sparse, accumulated by $p\times p$ blocks and solved by
one sparse LU factorisation reused for all output components. The local fit uses the eigendecomposition of
$G_k=\sum_{i\in I_k} a_k(x_i)\,\phi_k(x_i)\phi_k(x_i)^{\top}$ (eigenvalues below $10^{-10}$ of the largest discarded),
with $\lambda_\ell\in\{0,10^{-4},10^{-3},10^{-2},10^{-1},1,\infty\}$ chosen on validation; $\lambda_\ell=\infty$ reduces
\gscn{} to its trend. This classical partition-of-unity fit does not minimise the blended loss jointly, but it is
unaffected by the non-identifiability of heavily overlapping splats. In low-dimensional delay spaces
$\kappa\approx10$--16; in high dimension (Lorenz-96, $d\ge10$, $\kappa\ge45$) the global solve fills in and becomes
ill-posed.

\subsection{Complex-network guidance}\label{sec:networks}
Two networks propose starting points; neither decides the accuracy (\S\ref{sec:ablation};
\ssecref{sec:networks-full}). A $k$-nearest-neighbour recurrence network of the delay
vectors~\cite{marwan2009complex,donner2010recurrence} proposes the smallest embedding dimension $m_0$ at which neighbours
stay neighbours 20 steps ahead; since extra delays can average out float32 rounding, $m$ is then extended on validation,
keeping only embeddings stable in closed loop,
\begin{equation}
  \begin{aligned}
  m^\star &= \arg\min_{m\in\mathcal{S}} E(m),\\
  \mathcal{S}&=\Big\{m:\ 1-R^2(m)\le 1.5\,\min_{m'}\big(1-R^2(m')\big) + 10^{-7}\Big\}
  \end{aligned}
  \label{eq:msearch}
\end{equation}
($E$: a proxy field's validation MSE; $R^2$: its 30-step closed-loop validation $R^2$). The initial splats are allocated
over $C=128$ $k$-means cells, the nodes of a state-transition network~\cite{small2013complex}: cell $u$ gets
\begin{equation}
  n_u = \max\Big(1,\ \operatorname{round}\Big(K_0\,\frac{w_u}{\sum_v w_v}\Big)\Big)
  \label{eq:alloc}
\end{equation}
splats, with $K_0=192$ and $w_u=\sqrt{v_u}\,\big(1+\log(1+\varsigma_u)\big)$, so they concentrate where the attractor
is visited ($v_u$ visits) and where errors grow fastest ($\varsigma_u$: the spread of its successors over that of its
members); with a trend, splats go where it fails.

\subsection{Adaptive density control}\label{sec:densify}
After each fit every splat accumulates its blended squared residual and its opacity,
\begin{equation}
  e_k = \sum_{i\in I_k} a_k(x_i)\,\|F(x_i)-y_i\|^2,\quad o_k = \sum_{i\in I_k} a_k(x_i);
  \label{eq:accum}
\end{equation}
as in 3D Gaussian splatting~\cite{kerbl20233d}, near-transparent splats are pruned and those with the largest error
density $e_k/o_k$ split, while validation improves. Splat $k$ is kept if
$o_k>\max\big(Q_{0.02}(o),\ \min(3,\ 0.1\operatorname{median}o)\big)$ ($Q_{0.02}$: the 2\% quantile). Of the kept
splats, the $\lfloor K/4\rfloor$ with the largest $e_k$ are split if $e_k/o_k > f\cdot\operatorname{median}_{j}(e_j/o_j)$,
focus $f\in\{10,4\}$, into $\mu_k\pm\tfrac12\sigma_k\,\xi_k$ ($\xi_k\in\{-1,1\}^m$ random signs); widths are recomputed
and coefficients refitted. The cycle stops after 30 rounds, at $K\ge6000$, or after 15 rounds without a lower
validation MSE, and the best round is restored.

\subsection{Model selection}\label{sec:selection}
Every combination of trend (none or the best polynomial) and solver runs the densification schedule with $f=10$;
the best is rerun with $f=4$; candidates unstable in closed loop (Eq.~\eqref{eq:msearch}) are dropped, and the rest,
the trend alone included, compete on one-step validation MSE. Forecasts are clamped
to the training range widened by half its span, and an outlier guard, kept only if it lowers the validation MSE, falls
back on the median local prediction where the four local models disagree (\ssecref{sec:safeguards}).

\subsection{Full mode: hinged splats, members and a network referee}\label{sec:v4}

\begin{figure*}[t]
\centering
\includegraphics[width=\textwidth]{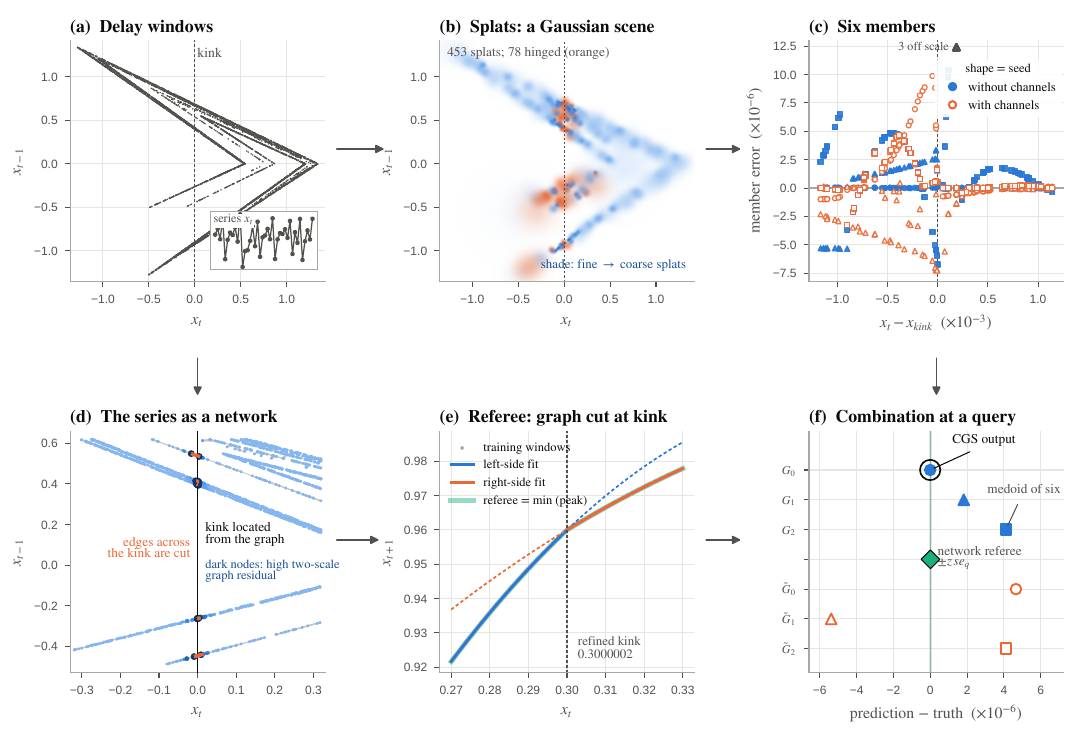}
\caption{\gscn{} in full mode at a glance (development data, initial condition 0, $n=10^4$). (a)~Lozi delay windows (two
values; kink along $x_t=0$). (b)~The fitted splat field on the same windows, rendered as alpha-composited Gaussians as
in 3D Gaussian splatting: 453 splats, fine (dark) to coarse (light); the 78 in orange carry hinges. (c)~Bent logistic
map (kink at $x_t=0.3$): near the kink the test errors of the six members split ($G_s$ without and $\tilde G_s$ with colour channels; marker shape: seed $s$).
(d)~The $k$-nearest-neighbour graph of the training windows, shaded by the two-scale residual $|4r(h)-r(2h)|$, which
locates the kink; the edges across it are cut. (e)~One-sided heat-kernel-weighted local quadratics meet at the refined
kink ($0.3000002$; true value $0.3$). (f)~At a disputed test point the medoid of the six would output a wrong member
($\sim\!4\times10^{-6}$ off); the referee's interval $\pm z\,se_x$ selects the member at the floor.}
\label{fig:algorithm}
\end{figure*}

Compact mode leaves its largest errors in a thin band around the kinks (\S\ref{sec:heldout}), and its single fit
depends on the seed and the floating-point environment. Full mode gives the splats hinges, fits several splat fields and
combines them, deferring to a network referee where they disagree (Fig.~\ref{fig:algorithm}).

\paragraph{Hinged splats} The local quadratic of Eq.~\eqref{eq:local} may carry $H_k\le2$ hinges,
\begin{equation}
  \tilde P_k(u)=P_k(u)+\sum_{j=1}^{H_k}\kappa_{kj}\,\psi_{kj}\big(n_{kj}^{\top}u-c_{kj}\big),
  \label{eq:kinked}
\end{equation}
kept only if they lower the validation error by more than 20\%; $F^{(m,s)}$ is Eq.~\eqref{eq:field} with these splats,
at window $m$ and seed $s$. Here $\psi_{kj}(s)=|s|$ where the splat's training points lie on both sides of the
hinge, $\pm s$ (their side) otherwise, so a splat never bends where it saw no data. The $\kappa_{kj}$ are fitted with
the $c_k$ by the same least squares; the unit normal $n_{kj}$ and offset $c_{kj}$ are searched on the training points
around the splat or taken from a neighbouring splat's line. The 20\% is a parsimony margin (cross-validation on the
training pairs when validation holds too few kink points).

\paragraph{Window composites and seeds} With $W=\{m^\ast-1,m^\ast,m^\ast+1\}$ around the selected window ($\{1,2,3\}$
when $m^\ast=1$), each cell $r$ of seed 0's splat centres takes the window $w_r$ that validates best on its own points;
the composite of seed $s$ is this gated model or the median over $W$, whichever validates better,
\begin{equation}
  \begin{aligned}
  G_s(x)&=F^{(w_{r(x)},s)}(x)\quad\text{or}\quad\operatorname*{med}_{m\in W}F^{(m,s)}(x),\\
  \bar G(x)&=\operatorname*{med}_{s\in\{0,1,2\}}G_s(x).
  \end{aligned}
  \label{eq:composite}
\end{equation}

\paragraph{Kink gate} Hinges are searched only where the training data look kinked: the 99.9th percentile of the
leave-one-out misfit of local quadratics must exceed 100 times its median. The misfit uses 60 nearest neighbours and windows 1--3.
On development data the ratio is 8--9 for series~3 and at least 164 for every kinked map.
The long-window path for finely sampled flows (R\"ossler) is in \ssecref{app:v4spec}.

\paragraph{Colour channels} On a kinked map each window model becomes the median of three forecasters, as a splat in
graphics carries several colour channels,
\begin{equation}
  \tilde F^{(m,s)}=\operatorname{med}\big\{F^{(m,s)},\,F_2^{(m,s)},\,F_4^{(m,s)}\big\},
  \label{eq:channels}
\end{equation}
where the splats of $F_h$ are placed by fitting the next $h$ values, then refitted on the one-step target and hinged:
the multi-step targets constrain where a splat puts its kink, as several views constrain a splat in 3D. $\tilde G_s$ is
their composite.

\paragraph{Six members and a network referee} On a kinked map \gscn{} keeps six members,
$(p_1,\dots,p_6)=(G_0,G_1,G_2,\tilde G_0,\tilde G_1,\tilde G_2)$. Near a kink they can disagree by orders of magnitude,
and wrong-branch members can agree with each other, so a vote can fail. Where they agree the output is their medoid
$\bar p$; where they split, a referee $r(x)$ decides: a heat-kernel-weighted local quadratic on neighbouring training
windows, fitted on the query's side of any kink located in the windows' $k$-nearest-neighbour graph, with standard
error $se_x$. A two-scale residual on the graph, the difference of local linear fits at bandwidths $h$ and
$2h$, cancels smooth curvature (order $h^2$) and keeps a kink (order $h$); grouped high-score nodes give kink
hyperplanes from the training series alone, at the members' window. $se_x$ comes from the referee's residual variance
and leverage (infinite where the neighbourhood leaves a direction undetermined). Near a located kink the referee uses
only points beyond a guard band set by the uncertainty of the kink's position, which the continuity of the two
one-sided fits refines. The medoid uses a fixed tie-breaking rule, so the result does not depend on the linear-algebra
library. With $z=\sqrt{2\ln n}$ (about the largest residual, in standard errors, among $n$ training windows),
$C=\{i:|p_i-r(x)|\le z\,se_x\}$ the members consistent with the referee and $se_{\min}$ its best precision on training
windows, \gscn{} predicts
\begin{equation}
  F_{\mathrm{CS}}(x)=\begin{cases}\bar p,&\max_ip_i-\min_ip_i\le z\,se_{\min}\,\text{or}\,|C|=6,\\
  p_{i^\ast},&0<|C|<6,\\ r(x),&C=\emptyset,\end{cases}
  \label{eq:v4}
\end{equation}
with $i^\ast=\arg\min_i|p_i-r(x)|$ ($\bar p$ where $r(x)$ is undefined). The last case covers queries in gaps
of the attractor, where every splat model extrapolates. The referee keeps the training windows, as the local baselines
do. On the long-window path (R\"ossler) full mode predicts $F_{\mathrm{long}}$, and on smooth maps $\bar G$, whose splats
carry no hinges. Every component was developed on initial conditions 0 and 1 of the
kinked maps and kept only if it lost no win there; \ssecref{app:v4spec} gives every setting.

%% ====================================================================
\section{Experimental protocol}\label{sec:protocol}

Trajectories are computed in 80-bit extended precision and cast to float32, so observations are exact up to float32
rounding, following Valle and Bruno~\cite{valle2025forecasting} (series and protocols: Table~\sref{tab:datasets}).
Every model starts from the same float32 windows. Scalar series are forecast recursively from 3{,}000 evenly spaced test windows and scored by the per-step and cumulative
$R^2$; full-state systems by the valid prediction time
(VPT). The cumulative $R^2$ pools every prediction of steps $1,\dots,h$:
$R^2_{\mathrm{cum}}(h) = 1 - \sum_{w}\sum_{j\le h}(x_{w,j}-\hat x_{w,j})^2 / \sum_{w}\sum_{j\le h}(x_{w,j}-\bar x_h)^2$,
with $\bar x_h$ the mean of the true values over the same set. The VPT is the first time, in Lyapunov times, at which the closed-loop normalised error
$\|\hat x(t)-x(t)\|/\sigma$ exceeds $\epsilon=0.5$, averaged over 40 well-separated test trajectories (on Lorenz-63
after a 50-step warm-up on the true states); test windows span 40 (Lorenz-63) to 120 (Lorenz-96) Lyapunov times, so no
forecast is truncated. Forecasts are fed back in float32, so the data's rounding, about $10^{-7}$ of the spread, reaches
the threshold after roughly $\ln(0.5/10^{-7})\approx15$ Lyapunov times.

\subsection{Baselines and how they were tuned}\label{sec:baselines}
The local baselines keep the training series and fit around every query: the \emph{Farmer--Sidorowich
predictor}~\cite{farmer1987predicting}, a polynomial of degree 1 or 2 on the $k$ nearest training windows, and the
\emph{S-map}~\cite{sugihara1994nonlinear}, a linear model with weights decaying with distance. The \emph{global
polynomial}, the model class of PolyProp~\cite{polyprop2025} and NG-RC~\cite{gauthier2021next} and \gscn{}'s own trend,
is fitted in double precision with a validated ridge; the others are an RBF network, NG-RC, an echo state network and,
on Santa Fe, an LSTM (Table~\sref{tab:baselines}). Every baseline sees the same data, uses every training point, is
tuned on the same validation segment by the same criterion, with search ranges extended past any edge, and is solved
stably. The local methods are run on \gscn{}'s window and on a window each selects by \gscn{}'s rule
(\S\ref{sec:selection}); the global polynomial selects its own; NG-RC also gets the closed-loop selection of
\ssecref{sec:closedloop} where reported. Unregularised, the double-precision fit and PolyProp's 512-bit fit give the
same polynomial of each degree in exact arithmetic, and with float32 data refitting in 512 bits changes no result
(\S\S\ref{sec:maps} and~\ref{sec:smooth}). A search whose selection falls on an end of its range is extended until it
is interior or reaches a limit of the method or a stated bound. Every least-squares problem is solved by blocked QR and
one SVD of the triangular factor, not by normal equations, whose solution near the rounding floor depends on the thread
count; the local methods fit each neighbourhood by SVD in centred, scaled coordinates with a validated ridge
(\ssecref{sec:baselines-full}). On Santa Fe an LSTM and an ESN with fixed architectures, and on dysts the published N-BEATS,
Transformer, TCN and neural ODE, serve as references. SINDy~\cite{brunton2016discovering} fits a sparse member of the
global polynomial class and Chronos~\cite{zhang2025zeroshot,ansari2024chronos} quantises the series far more coarsely
than float32, so neither is run.

%% ====================================================================
\section{Results}\label{sec:results}

Benchmark runs use compact mode; full mode is tested on fresh data in \S\ref{sec:fresh}.\footnote{Validation kept
the splats on every series observed through delays and none on the smooth full-state flows (Table~\sref{tab:regimes}).}
\ifjournal\else %% SUPP-BEGIN Selected configurations

\fi %% SUPP-END

\subsection{Chaotic maps}\label{sec:maps}

\begin{ptable}[tbp]{tab_maps}\centering
\caption{The Valle and Bruno benchmark: one-step test MSE of compact \gscn{}, of the rounding floor (a forecast equal to the exact, unrounded next value, scored against the float32 targets: the precision of the data), and of the best of the other methods run here (Table~\sref{tab:maps-baselines}); and the variance left unexplained over 20 closed-loop steps, $1-R^2_{\mathrm{cum}}(20)$, from 3{,}000 test windows, for compact \gscn{} and the best other method ($9\times10^5$ training, $5\times10^4$ validation and $5\times10^4$ test points per series; float32 feedback). FS: Farmer--Sidorowich predictor; $^\ast$: on its own window.}\label{tab:maps}
\small
\fitwidth{%
\begin{tabular}{@{}lcccccc@{}}
\toprule
Series & $\lambda$ & \gscn{} & \begin{tabular}[b]{@{}c@{}}Rounding\\floor\end{tabular} & \begin{tabular}[b]{@{}c@{}}Best other\\method\end{tabular} & \begin{tabular}[b]{@{}c@{}}$1-R^2_{\mathrm{cum}}(20)$\\\gscn{}\end{tabular} & \begin{tabular}[b]{@{}c@{}}$1-R^2_{\mathrm{cum}}(20)$\\best other\end{tabular}\\
\midrule
1. Logistic ($\mu=3.854$) & 0.019 & $1.79\times10^{-16}$ & $1.70\times10^{-16}$ & $2.08\times10^{-16}$ (ESN) & $1.4\times10^{-15}$ & $7.8\times10^{-15}$ (ESN)\\
2. Logistic ($\mu=3.874$) & 0.456 & $4.34\times10^{-16}$ & $1.97\times10^{-16}$ & $4.27\times10^{-16}$ (FS$^\ast$) & $1.9\times10^{-7}$ & $1.6\times10^{-7}$ (FS$^\ast$)\\
3. Logistic ($\mu=4$) & 0.693 & $4.07\times10^{-16}$ & $1.63\times10^{-16}$ & $4.03\times10^{-16}$ (FS) & $6.8\times10^{-5}$ & $6.6\times10^{-5}$ (FS)\\
4. Logistic--tent ($\mu=3.999$, $\Gamma=1.323$) & 0.856 & $1.67\times10^{-15}$ & $1.66\times10^{-16}$ & $1.61\times10^{-15}$ (S-map) & $4.2\times10^{-2}$ & $5.2\times10^{-2}$ (FS)\\
5. Logistic--tent ($\mu=3.999$, $\Gamma=1.686$) & 1.051 & $1.85\times10^{-14}$ & $1.94\times10^{-16}$ & $4.00\times10^{-15}$ (S-map) & $3.9\times10^{-1}$ & $3.9\times10^{-1}$ (FS)\\
6. Logistic--tent ($\mu=3.999$, $\Gamma=1.989$) & 1.253 & $6.13\times10^{-15}$ & $1.87\times10^{-16}$ & $7.32\times10^{-15}$ (S-map) & $7.0\times10^{-1}$ & $7.0\times10^{-1}$ (S-map)\\
\bottomrule
\end{tabular}}
\end{ptable}

Table~\ref{tab:maps} gives compact \gscn{}'s one-step error beside the rounding floor and the best other method. The global models come within a factor of \num{smooth.globalfloor.max} of the floor on
series 1--3 but are $\numm{m4.ratioglobal}$, $\numm{m5.ratioglobal}$ and $\numm{m6.ratioglobal}$ times less accurate
than \gscn{} on series 4--6, at every degree tried~\cite{bernstein1914}, in double or 512-bit
precision. The global models are the validated polynomial, a Chebyshev polynomial of degree up to 256, NG-RC and the ESN
(Table~\sref{tab:maps-baselines}); a 512-bit monomial fit matches the double-precision one to $\numm{pp.map.dev}$ up to
degree \num{pp.map.deg}. Given every training point, the Farmer--Sidorowich predictor and the S-map are as accurate as
compact \gscn{}, but keep the
whole series ($9\times10^5$ values), where compact \gscn{} stores $\numm{m4.cs.size}$ numbers on series 4
(\S\ref{sec:size}). Their one-step errors are within 25\% of compact \gscn{}'s on every series but series 5 (S-map lower), most differences
unresolved by paired bootstrap intervals; in closed loop compact \gscn{}'s lead on series 4 is resolved (the best local
method leaves \num{ci.4.ratio20} times its unexplained variance). Run until their cumulative $R^2$ falls to zero,
compact \gscn{}'s forecasts keep skill within \num{lr2.cslocal.maxgap} steps of the local methods' on every series,
while on the kinked series the global models lose the state within a few steps (\ssecref{sec:longr2}). Full mode keeps skill in closed loop as long as the exact map or longer on every map but one
(Table~\ref{tab:full-horizons}). Learning curves (Table~\sref{tab:learning}): with $10^3$, $10^4$ and $10^5$
training points the S-map is often the most accurate one step ahead with little data, while compact \gscn{} leaves the
least 20-step variance on series 4--6 at every size. Full mode, fitted afterwards with the frozen code (development
data, not a test), has the lowest one-step error on \num{full.learn.mse.wins.1000}, \num{full.learn.mse.wins.10000} and
\num{full.learn.mse.wins.100000} of the six series and is within a factor of \num{full.learn.mse.worst} of the lowest on
the others; on series 4--6 with $10^3$ points the best other model's error is $\numm{full.learn.mse.kgain.min.1000}$ to
$\numm{full.learn.mse.kgain.max.1000}$ times full mode's, which is within a factor of \num{full.ratio.k.max.1000} of the
exact map's. On series 3 its 20-step error is up to \num{full.learn.unexpl.worst} times compact mode's.

\begin{ptable}[tbp]{tab_full_horizons}\centering
\caption{\gscn{} in full mode on all the paper's scalar series, fitted with the code frozen for the confirmation on the first $n$ training points of each (validation on the next $\max(250,n/4)$, the full test segment): one-step MSE relative to the exact map's, both over the first step of the 3{,}000 test windows, and the step at which the cumulative $R^2$ of the closed-loop forecast over those windows falls to 0, for \gscn{} / the exact map applied to the float32 observations (float32 feedback; series 1: 500 windows), printed once where the two coincide. Where it did not fall to 0 within the cap: the cap and, in parentheses, the cumulative $R^2$ there; on series 1 no forecast leaves the bands of its attractor, so none falls to 0. The flows have no exact map. A reported analysis on development data, not a pre-registered test. --: not run.}\label{tab:full-horizons}
\small
\fitwidth{%
\begin{tabular}{@{}lcccccc@{}}
\toprule
Series & \multicolumn{2}{c}{$n=10^3$} & \multicolumn{2}{c}{$n=10^4$} & \multicolumn{2}{c}{$n=10^5$}\\
\cmidrule(lr){2-3}\cmidrule(lr){4-5}\cmidrule(l){6-7}
 & MSE / exact & horizon & MSE / exact & horizon & MSE / exact & horizon\\
\midrule
1. Logistic ($\mu=3.854$, $\lambda=0.019$) & 0.17 & $>$20{,}000 (1.00) & 0.07 & $>$20{,}000 (1.00) & 0.04 & $>$20{,}000 (1.00)\\
2. Logistic ($\mu=3.874$, $\lambda=0.456$) & 0.37 & 76 / 70 & 0.33 & 76 / 70 & 0.32 & 76 / 70\\
3. Logistic ($\mu=4$, $\lambda=0.693$) & 0.35 & 50 / 48 & 0.32 & 51 / 48 & 0.30 & 51 / 48\\
4. Logistic--tent ($\mu=3.999$, $\Gamma=1.323$, $\lambda=0.856$) & 1.07 & 39 & 0.83 & 39 & 0.54 & 41 / 39\\
5. Logistic--tent ($\mu=3.999$, $\Gamma=1.686$, $\lambda=1.051$) & 0.76 & 32 / 31 & 0.94 & 32 / 31 & 0.51 & 33 / 31\\
6. Logistic--tent ($\mu=3.999$, $\Gamma=1.989$, $\lambda=1.253$) & 1.19 & 26 & 0.85 & 27 / 26 & 0.58 & 27 / 26\\
H\'enon & -- & -- & 0.20 & 87 / 78 & 0.15 & 89 / 78\\
Skew tent & 1.18 & 56 & 1.01 & 55 / 56 & 0.72 & 56\\
Lozi & 1.05 & 76 / 74 & 0.85 & 75 / 74 & 0.72 & 75 / 74\\
Border collision & 1.53 & 85 & 0.90 & 86 / 85 & 0.57 & 90 / 85\\
Quadratic tent & 1.23 & 57 & 1.00 & 57 & 0.41 & 60 / 57\\
Curved skew tent & 1.15 & 68 & 1.00 & 68 & 0.90 & 68\\
Bent logistic & 3.09 & 47 & 0.68 & 49 / 47 & 0.41 & 51 / 47\\
Lorenz-63, $x$ & -- & -- & -- & 1542 & -- & 3368\\
R\"ossler, $x$ & -- & -- & -- & $>$25{,}000 (0.61) & -- & $>$25{,}000 (0.62)\\
\bottomrule
\end{tabular}}
\end{ptable}

\subsection{Held-out kinked maps: a pre-registered test}\label{sec:heldout}

This test is where compact mode reached its limit, and the limit is what full mode was built to remove; we report it in
that order: the question, the failure, the change of method, and the confirmation of the changed method.

\paragraph{The question} A protocol committed before any method was run tested compact \gscn{} on three further kinked maps, piecewise linear
and so favourable to local linear fits: the skew tent map, the Lozi map~\cite{lozi1978attracteur} and the
border-collision normal form~\cite{banerjee1998robust}. The parameters are: skew tent $a=0.7$; Lozi $A=1.7$, $B=0.5$;
border-collision normal form $(t,d)=(1.4,0.2)$ for $x\le0$, $(-1.6,0.3)$ for $x>0$, $\mu=1$, the last two observed
through $x$ (largest Lyapunov exponents 0.61, 0.47, 0.40). Every method is trained on the first $n=10^3$ to
$9\times10^5$ points, validated on the next $\max(250,n/4)$ and scored on the same 3{,}000 test windows. H1: at
$n=10^3$ and $10^4$ \gscn{} has the lowest $1-R^2_{\mathrm{cum}}(20)$ of all fitted methods in at least four of the six
map--size cells. H2: at $n=9\times10^5$ its one-step MSE is within a factor of three of the best fitted method's and of
the exact map's, on every map. H3: at $n=9\times10^5$ the best global model (global polynomial, NG-RC) has a one-step
MSE at least $10^3$ times \gscn{}'s, on every map.

\paragraph{The limitation} \emph{H1 failed}: \gscn{} had the lowest closed-loop error in \num{ho.H1.wins} of \num{ho.H1.cells} cells, both on the skew tent. \emph{H2 failed} on the border-collision map. \emph{H3 held}: the
best global model was $\numm{ho.H3.min}$ to $\numm{ho.H3.max}$ times less accurate (Table~\ref{tab:heldout}, compact
mode's verdicts). On Lozi and the border-collision map a kink is a line in delay space, which unhinged splats resolve
only with many points.

\begin{ptable}[tbp]{tab_heldout}\centering
\caption{Held-out kinked maps, pre-registered (HELDOUT\_PROTOCOL.md, committed before any method was run): training on the first $n$ points, validation on the next $\max(250, n/4)$, the same 3{,}000 test windows at every size. One-step test MSE and $1-R^2_{\mathrm{cum}}(20)$ of compact \gscn{} and of the best of the eight other fitted methods (in parentheses: FS Farmer--Sidorowich, S-map, $^\ast$ on its own window, RBF network, poly. global polynomial, NG-RC, ESN; the ESN up to $n=10^5$). Full: \gscn{} in full mode, fitted after the test with the code frozen for the confirmation, on the same points and scored by the same code, up to $n=10^5$ (--: not run); it is not part of the pre-registered test. Bold: the lowest.}\label{tab:heldout}
\small
\fitwidth{%
\begin{tabular}{@{}cccccccc@{}}
\toprule
Map & $n$ & \multicolumn{3}{c}{one-step MSE} & \multicolumn{3}{c}{$1-R^2_{\mathrm{cum}}(20)$}\\
\cmidrule(lr){3-5}\cmidrule(l){6-8}
 & & compact & full & best other & compact & full & best other\\
\midrule
Skew tent & $10^{3}$ & $1.1\times10^{-11}$ & \textbf{$1.7\times10^{-15}$} & $2.6\times10^{-9}$ (FS) & $3.3\times10^{-3}$ & \textbf{$1.6\times10^{-3}$} & $1.3\times10^{-2}$ (S-map)\\
 & $10^{4}$ & $1.1\times10^{-13}$ & \textbf{$1.5\times10^{-15}$} & $2.7\times10^{-12}$ (FS) & $1.5\times10^{-3}$ & \textbf{$1.5\times10^{-3}$} & $2.1\times10^{-3}$ (FS)\\
 & $10^{5}$ & $1.7\times10^{-15}$ & \textbf{$1.1\times10^{-15}$} & $2.2\times10^{-15}$ (FS) & $1.5\times10^{-3}$ & \textbf{$9.9\times10^{-4}$} & $2.1\times10^{-3}$ (FS)\\
 & $9\times10^{5}$ & \textbf{$1.2\times10^{-15}$} & -- & $1.7\times10^{-15}$ (FS) & \textbf{$1.3\times10^{-3}$} & -- & $1.6\times10^{-3}$ (S-map)\\
Lozi & $10^{3}$ & $1.3\times10^{-5}$ & \textbf{$1.3\times10^{-15}$} & $3.2\times10^{-6}$ (ESN) & $3.3\times10^{-1}$ & \textbf{$1.0\times10^{-7}$} & $8.8\times10^{-2}$ (S-map)\\
 & $10^{4}$ & $1.7\times10^{-7}$ & \textbf{$1.1\times10^{-15}$} & $2.1\times10^{-8}$ (FS) & $4.7\times10^{-2}$ & \textbf{$8.7\times10^{-8}$} & $1.1\times10^{-2}$ (S-map)\\
 & $10^{5}$ & $9.0\times10^{-10}$ & \textbf{$9.2\times10^{-16}$} & $2.2\times10^{-11}$ (S-map) & $8.7\times10^{-3}$ & \textbf{$7.9\times10^{-8}$} & $1.9\times10^{-3}$ (S-map$^\ast$)\\
 & $9\times10^{5}$ & $3.6\times10^{-15}$ & -- & \textbf{$2.8\times10^{-15}$} (S-map) & \textbf{$3.0\times10^{-4}$} & -- & $3.2\times10^{-4}$ (S-map)\\
Border collision & $10^{3}$ & $6.2\times10^{-7}$ & \textbf{$7.2\times10^{-16}$} & $8.3\times10^{-8}$ (S-map$^\ast$) & $1.4\times10^{-1}$ & \textbf{$9.1\times10^{-9}$} & $2.8\times10^{-2}$ (S-map$^\ast$)\\
 & $10^{4}$ & $1.9\times10^{-8}$ & \textbf{$4.6\times10^{-16}$} & $5.4\times10^{-11}$ (S-map) & $8.9\times10^{-2}$ & \textbf{$5.8\times10^{-9}$} & $7.7\times10^{-4}$ (S-map)\\
 & $10^{5}$ & $1.3\times10^{-11}$ & \textbf{$2.6\times10^{-16}$} & $3.5\times10^{-14}$ (S-map) & $1.2\times10^{-4}$ & \textbf{$1.3\times10^{-9}$} & $1.2\times10^{-4}$ (FS)\\
 & $9\times10^{5}$ & $6.8\times10^{-14}$ & -- & \textbf{$6.2\times10^{-16}$} (FS$^\ast$) & $4.0\times10^{-6}$ & -- & \textbf{$1.1\times10^{-7}$} (S-map$^\ast$)\\
\bottomrule
\end{tabular}}
\end{ptable}

A second protocol tested scalar kinked maps with curved pieces, where a local linear fit is not exact: H1$'$
\num{ho2.H1.verdict} (\num{ho2.H1.wins} of \num{ho2.H1.cells} cells), H2$'$ \num{ho2.H2.verdict}, H3$'$
\num{ho2.H3.verdict} ($\numm{ho2.H3.min}$ to $\numm{ho2.H3.max}$; Table~\sref{tab:heldout2}).\footnote{$x'=1-|x|-0.9x^2$,
$x'=(T+0.5T^2)/1.5$ with $T$ a skew tent map, and $x'=4B(1-B)$ with $B$ a broken line (largest Lyapunov exponents 0.61,
0.53, 0.69), with the same hypotheses; with all the data \gscn{}'s one-step error was \num{ho.quadtent.H2.vsbest},
\num{ho.curvedtent.H2.vsbest} and \num{ho.bentlogistic.H2.vsbest} times the best fitted method's.}

\paragraph{The change of method} Full mode (\S\ref{sec:v4}) answers the failure directly: its splats carry hinges that
represent a kink line exactly, its colour channels and seeds give several fits that fail in different places, and its
referee locates the kinks in the training windows' neighbour graph and decides where the fits disagree. It was built
after both tests, with these maps in its development data, so its results here (the column `full' of
Table~\ref{tab:heldout}) are not a test.

\paragraph{The confirmation} The test of the changed method is the pre-registered confirmation of \S\ref{sec:fresh}: full
mode was frozen before its data were generated and scored on fresh initial conditions of five of these six maps (all but the skew tent) and of five other systems, where it
beat both local methods on \num{f910.unitwins} of \num{f910.units} units.

\subsection{Computational cost}\label{sec:size}\label{sec:cost}
\begin{figure}[tbp]
\centering\includegraphics[width=\colfigw]{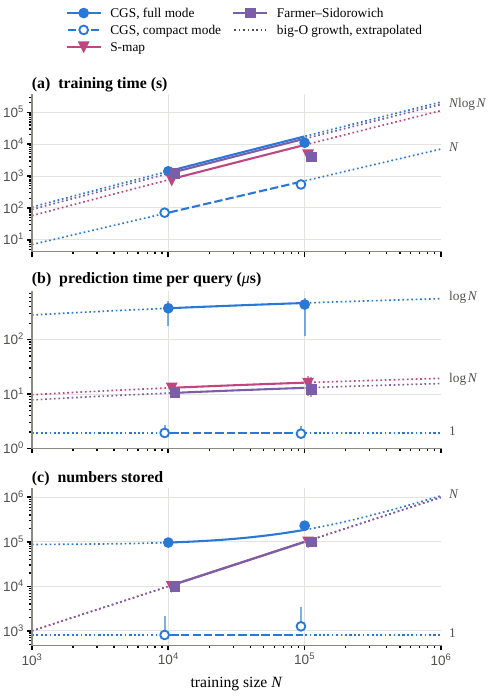}
\caption{Cost against the training size $N$ on the nine maps (both fresh initial conditions): (a)~training time,
(b)~prediction time per query, (c)~numbers stored, for \gscn{} in full and compact mode, the S-map and the Farmer--Sidorowich
predictor (each with every input view). Markers: medians, bars: interquartile ranges, measured in the confirmation's
jobs on the same machines (one thread). Lines: each method's growth in $N$ from Eq.~\eqref{eq:cost} (the other
parameters fixed), anchored at its median at $N=10^4$; dotted beyond the measured sizes. Compact storage is bounded
because the number of splats per field is capped (6\,000); full mode and the local methods keep the training series.}
\label{fig:scaling}
\end{figure}

With $N$ training points, window length $m$, $K$ splats of $p=(m+1)(m+2)/2$ coefficients and a trend of $D_q$
features (Table~\sref{tab:complexity} defines the rest), compact \gscn{} and a local method (Farmer--Sidorowich, S-map,
$k$ neighbours) cost
\begin{align}
T^{\mathrm{CGS}}_{\mathrm{train}}&=O\big(WCR\,N(\log K+n_bp^2)\big),\nonumber\\
t^{\mathrm{CGS}}_{\mathrm{pred}}&=O(\log K+n_bp+D_q),\nonumber\\
S^{\mathrm{CGS}}&=K(m+1+p)+D_q,\nonumber\\
T^{\mathrm{loc}}_{\mathrm{train}}&=O\big(N\log N+GVkp_\ell^2\big),\label{eq:cost}\\
t^{\mathrm{loc}}_{\mathrm{pred}}&=O(\log N+kp_\ell^2),\qquad S^{\mathrm{loc}}=N,\nonumber
\end{align}
with $T$ the training time (leading terms for $N\gg K$; the full expressions in Table~\sref{tab:complexity}), $t$ the time of one prediction and $S$ the numbers stored. \gscn{} moves the cost into
training: it is linear in $N$ for a given $K$, and the prediction and the storage depend on $K$, not on $N$, where a
local method keeps all $N$ points and fits a model at every query. Full mode adds the referee, $O(N(\log N+k_gm^2))$
to train and the $N$ training windows to store (\S\ref{sec:v4}). Figure~\ref{fig:scaling} sets these laws against the
times and sizes measured in the confirmation. All fitting is closed-form least squares, with no gradient descent; the only randomness is seeded.

\paragraph{Training time} Full \gscn{} trained in a median of \num{f910.time.1e4.cs} minutes at $n=10^4$ and
\num{f910.time.1e5.cs} hours at $10^5$ on one thread, against \num{f910.time.1e4.base} minutes and
\num{f910.time.1e5.base} hours for the two baselines with all their input views; its time grew as
$n^{\num{f910.scale.fit}}$, as Eq.~\eqref{eq:cost} implies, against $n^{\num{f910.scale.smfit}}$ and
$n^{\num{f910.scale.lpfit}}$ for the tuning of the S-map and the Farmer--Sidorowich predictor. Its three seeds are
independent fits, so run in parallel they would cut the wall-clock time by a factor of about
\num{f910.par.speedup.1e4}. Compact training on the held-out maps took \num{train.cs.min} to \num{train.cs.max} minutes
per map, against \num{train.local.min} to \num{train.local.max} for the local methods (Table~\sref{tab:cost}).

\paragraph{Memory and prediction} Compact \gscn{}'s number of splats grew only as about $N^{0.2}$--$N^{0.3}$ on the
held-out maps (\num{ho.lozi.1000.cs.splats} to \num{ho.lozi.900000.cs.splats} on Lozi from $10^3$ to $9\times10^5$
points), so with all the data a compact prediction on Lozi takes \num{cost.lozi.900000.cs.us}~$\mu$s, against
\num{cost.lozi.900000.lp.us} and \num{cost.lozi.900000.smap.us}~$\mu$s for the Farmer--Sidorowich predictor and the
S-map. At equal accuracy the memory budget is the point (\S\ref{sec:budget}): at $n=10^5$ compact \gscn{} needed
\num{bud.k1.100000.gm.min} to \num{bud.k1.100000.gm.max} times less memory than the cheapest local model of the same
validation error. Full mode pays for its accuracy: it keeps the training windows for its referee and
predicts in \num{f910.pred.1e4.cs}\,$\mu$s per query against \num{f910.pred.1e4.base}\,$\mu$s.

\ifjournal\else %% SUPP-BEGIN Measured cost on the held-out maps

\fi %% SUPP-END

\subsection{Memory and speed at equal accuracy: a repeated study}\label{sec:budget}

A study fixed in advance tested whether compact \gscn{} gives the Farmer--Sidorowich predictor's accuracy with less
memory and faster prediction. The study covers ten systems (series 3--6, five held-out kinked maps, R\"ossler), $n=10^3$ to
$9\times10^5$, five initial conditions and three training seeds each. \gscn{} is held to serialised budgets of 1~KiB to
1~MiB; the local methods get all training points and, at smaller budgets, subsets chosen four ways (first values,
random, coverage, error-driven), each tuned on validation. Prediction is timed in batches of 3\,000 windows on one
thread. Of \num{bud.runs.planned} planned runs, \num{bud.runs.done} completed.

At the target of 1.1 times the full predictor's validation error, on the six scalar kinked maps with $n=10^5$, \gscn{}
met it in \num{bud.k1.100000.val} of \num{bud.k1.100000.runs} runs and then needed \num{bud.k1.100000.gm.min} to
\num{bud.k1.100000.gm.max} times less memory and predicted \num{bud.k1.100000.gt.min} to \num{bud.k1.100000.gt.max}
times faster than the cheapest local model meeting it (\num{bud.k1.900000.gm.min}--\num{bud.k1.900000.gm.max} and
\num{bud.k1.900000.gt.min}--\num{bud.k1.900000.gt.max} with $9\times10^5$ points; every memory gain's interval above
one; Fig.~\sref{fig:budget-cost}, Table~\sref{tab:budget-targets}). On series 4 \gscn{} reaches the full local
methods' precision with a few KiB, where they need about a hundred (Fig.~\sref{fig:budget}); on Lozi and R\"ossler the
local methods are the more accurate at every budget. The target is tight: \gscn{}'s test error stays within it in
\num{bud.k1.100000.test} of the \num{bud.k1.100000.val} runs. On the border-collision map the memory gain is
\num{bud.bcnf.100000.gm}, on Lozi \gscn{} meets the target in no run at $10^5$, and with $10^3$ points the advantage
reverses (Table~\sref{tab:budget-small}).

\ifjournal\else %% SUPP-BEGIN Budget study: cost at the target

\fi %% SUPP-END

\paragraph{H\'enon map and partially observed flows} On the H\'enon map compact \gscn{}, the global polynomial and the
Farmer--Sidorowich predictor are within a factor of two of the rounding floor, and full mode keeps a positive cumulative
$R^2$ for \num{full.h.henon.10000} and \num{full.h.henon.100000} steps ($n=10^4$, $10^5$), longer than the exact map's
\num{full.hx.henon.10000}. With only $x$ of Lorenz-63 and R\"ossler observed, compact mode trails the
Farmer--Sidorowich predictor, which refits a quadratic around every query (full mode: \S\ref{sec:fresh},
Table~\ref{tab:full-horizons}).

\subsection{Experimental data: the Santa Fe laser}\label{sec:santafe}

On data set A of the Santa Fe competition~\cite{weigend1994time}, 1{,}000 noisy samples of a chaotic laser forecast
100 steps through a collapse of the oscillation, compact \gscn{} ($m=\num{sf.cs.m}$) and the S-map are level over 100
steps (NMSE \num{sf.GSCN.100} and \num{sf.SmapSugihara.100}), ahead of the Farmer--Sidorowich predictor
(\num{sf.FarmerSidorowichlocalpolynomial.100}) and the LSTM (\num{sf.LSTM.100}); full mode ($m=\num{full.sf.m}$) is
worse (NMSE \num{full.sf.100}).\footnote{The last 100 points are used for validation (Table~\sref{tab:santafe},
Fig.~\sref{fig:santafe}). The tuned ESN is best over 50 steps and then diverges; the global polynomial and NG-RC
diverge; all stay far behind the two best competition entries (0.027 and 0.077). Full mode's NMSE at 25 and 50 steps is
\num{full.sf.25} and \num{full.sf.50}, and at 100 steps no better than the mean. A single forecast carries no error bar;
the comparison is indicative.}

\ifjournal\else %% SUPP-BEGIN Santa Fe laser: table

\fi %% SUPP-END

\subsection{Observation noise}\label{sec:noise}
With Gaussian observation noise of $\varepsilon=10^{-6}$ to $10^{-2}$ times the standard deviation, every one-step error
grows as $\varepsilon^2$ and compact \gscn{}'s is \num{noise.ratio.min} to \num{noise.ratio.max} times the better local
method's; full mode keeps the lowest error on the kinked series in \num{noise.full.wins} of \num{noise.full.cells}
cases, but its lead shrinks from \num{noise.full.gain0.min}--\num{noise.full.gain0.max}$\times$ to at most
\num{noise.full.gainhi.max}\% from $\varepsilon=10^{-4}$. The noise study uses series 2 and 4--6, $10^5$ training points, each model
selected on the noisy validation data and scored against the clean series (Fig.~\sref{fig:noise},
Table~\sref{tab:noise-full}, \ssecref{sec:noise-full}); validation gives \gscn{} longer windows and fewer splats as the
noise grows. Already at $\varepsilon=10^{-6}$ the best one-step error on the kinked maps is
\num{noise.low.min}--\num{noise.low.max} times its noise-free value: the precision that separates \gscn{} and the local
methods from the global models is a property of clean data.

\ifjournal\else %% SUPP-BEGIN Observation noise: full mode

\fi %% SUPP-END

\begin{figure*}[tbp]
\centering\includegraphics[width=0.92\textwidth]{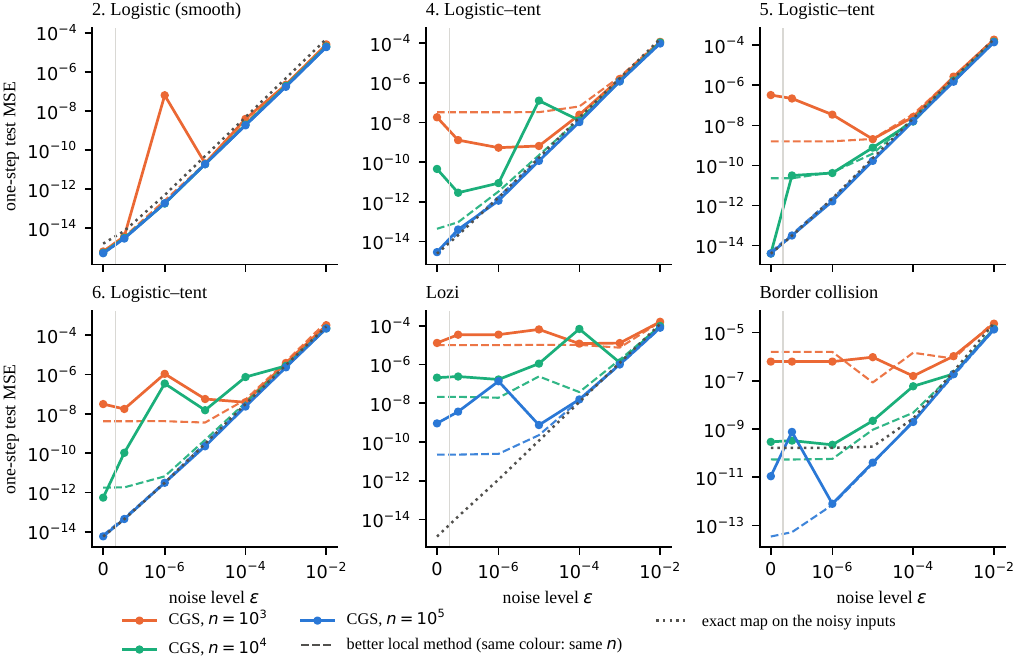}
\caption{Where noise takes over: one-step test MSE against the noise level $\varepsilon$ (relative to the series'
standard deviation; $\varepsilon=0$ at the left edge), for compact \gscn{} (solid) and the better of the local polynomial
and the S-map (dashed) at three training sizes, and the exact map applied to the noisy inputs (dotted). Each point is
one job of the pre-registered study (\nolinkurl{CSF_NOISE_PROTOCOL.md}): one noise realisation, 3{,}000 test windows.}
\label{fig:noise-transition}
\end{figure*}

\paragraph{Where noise takes over} A second study, committed with its hypotheses before any of its
\num{nt.jobs} jobs ran (\nolinkurl{CSF_NOISE_PROTOCOL.md}), asked where the error stops being set by the approximation and
starts being set by noise, and how that depends on the amount of data and on the kink. It crossed six systems (series 2
as the smooth control; series 4--6, kinked at $x=0.5$; the Lozi map and the border-collision normal form, kinked on the
line $x_t=0$ in delay space), three training sizes ($n=10^3$, $10^4$, $10^5$) and seven noise levels
($\varepsilon=0$ and $10^{-7}$ to $10^{-2}$), with compact \gscn{}, the two local methods and the exact map on the noisy
inputs, each model selected on noisy validation data and scored against the clean series (Fig.~\ref{fig:noise-transition}).
The \emph{transition level} $\varepsilon^\ast$ is the smallest noise level at which a method's one-step error is more
than twice its noise-free error (Table~\ref{tab:noise-transition}).

\begin{table}[tbp]
\centering
\caption{Transition level $\varepsilon^\ast$, the smallest noise level of the grid at which the one-step MSE exceeds
twice its noise-free value, for compact \gscn{} (CGS) and the better local method (local); --: no level does.}
\label{tab:noise-transition}
\small\fitwidth{% written by experiments/csf_noise/analyse_noise.py
\begin{tabular}{@{}lcccccc@{}}
\toprule
 & \multicolumn{2}{c}{$n=10^3$} & \multicolumn{2}{c}{$n=10^4$} & \multicolumn{2}{c}{$n=10^5$}\\
\cmidrule(lr){2-3}\cmidrule(lr){4-5}\cmidrule(l){6-7}
System & CGS & local & CGS & local & CGS & local\\
\midrule
2. Logistic (smooth) & $10^{-7}$ & $10^{-7}$ & $10^{-7}$ & $10^{-7}$ & $10^{-7}$ & $10^{-7}$\\
4. Logistic--tent & $10^{-3}$ & $10^{-3}$ & $10^{-5}$ & $10^{-7}$ & $10^{-7}$ & $10^{-7}$\\
5. Logistic--tent & $10^{-3}$ & $10^{-4}$ & $10^{-7}$ & $10^{-5}$ & $10^{-7}$ & $10^{-7}$\\
6. Logistic--tent & $10^{-6}$ & $10^{-4}$ & $10^{-7}$ & $10^{-6}$ & $10^{-7}$ & $10^{-7}$\\
Lozi & $10^{-7}$ & $10^{-2}$ & $10^{-5}$ & $10^{-5}$ & $10^{-7}$ & $10^{-5}$\\
Border collision & $10^{-2}$ & $10^{-2}$ & $10^{-5}$ & $10^{-5}$ & $10^{-7}$ & $10^{-6}$\\
\bottomrule
\end{tabular}
}
\end{table}

More data move the transition to lower noise, as the second hypothesis predicted (it \num{nt.h2.verdict} on all
\num{nt.h2.ok} kinked systems): with $10^5$ points compact \gscn{}'s noise-free error is so small (at the rounding floor
on series 2--6) that noise of $\varepsilon=\num{nt.h2.big}$ already doubles it on every system, whereas with $10^3$ points
the approximation error is larger and noise first doubles it at $\varepsilon=\num{nt.h2.small.min}$ to
$\num{nt.h2.small.max}$, depending on the system. Once noise dominates, the methods converge: at
$\varepsilon\ge10^{-3}$ compact \gscn{}'s error was within a factor of 1.5 of the better local method's in
\num{nt.h1.within} of \num{nt.h1.cells} cells (ratios \num{nt.h1.min}--\num{nt.h1.max}), so the first hypothesis
\num{nt.h1.verdict}, on one cell (\num{nt.h1.fail.system}, $n=\num{nt.h1.fail.n}$, $\varepsilon=\num{nt.h1.fail.eps}$,
ratio \num{nt.h1.fail.ratio}); at $\varepsilon\ge10^{-3}$ with $10^4$ or more points compact \gscn{} stayed below the
exact map applied to the noisy inputs in \num{nt.below.exact} of \num{nt.below.cells} cells, because a window of past values averages some of the
noise away. The third hypothesis \num{nt.h3.verdict}, and its failure is informative: without noise at $n=10^4$, the
better local method's error sat nearer the kink than compact \gscn{}'s on only \num{nt.h3.part1} of the five kinked
systems, because compact mode, which has no hinges, also leaves most of its error at the kink where the kink is hard
to resolve (\num{nt.h3.cs.series4}\% within the 1\% of test points nearest it on series 4, \num{nt.h3.cs.bcnf}\% on the
border-collision map, against \num{nt.h3.base.bcnf}\% for the local method there); removing that error is what full
mode's hinges and referee are for (\S\ref{sec:heldout}). At $\varepsilon=10^{-3}$ the error spreads over the attractor:
every method kept at most \num{nt.h3.others.max}\% of it near the kink, except the S-map on the border-collision map
(\num{nt.h3.over.share}\%). With $10^3$ points compact \gscn{} is often less accurate than the local methods, and a
selection on the 250 validation points can fail outright (series 2 at $\varepsilon=10^{-6}$), as the learning curves
already showed (\S\ref{sec:maps}).

\subsection{Ablation: what each component of compact mode contributes}\label{sec:ablation}
Refitting compact \gscn{} without each component (Table~\sref{tab:abl-components}) shows that the accuracy on the
kinked maps comes from the splats and their density control: without it series 4--6 stay at
$\numm{abl.kinked.nodensify.min}$ to $\numm{abl.kinked.nodensify.max}$, and the trend alone diverges in closed loop; on
series 2, 3 and H\'enon every variant is within \num{abl.smooth.spread}\% of the complete model. The complex-network
components do not decide the accuracy. The ablation uses series 2--6 and H\'enon, $2\times10^5$ training points, the benchmark's
window; variants: no splats (trend alone), no density control (initial splats fitted once), no trend, uniform or
$k$-means placement. Placement moves the kinked errors by up to two orders of magnitude with no consistent winner
(network placement best on series 4, worst on 5 and 6), within the selection effects of one run; the recurrence network
proposes $m=1$ on every map (Table~\sref{tab:abl-dims}; false nearest neighbours and Cao's method mostly fail on these
non-invertible maps), only a start for the validation search.

\subsection{Smooth, fully observed systems: the polynomial limit}\label{sec:smooth}

On full-state Lorenz-63 and Lorenz-96 the one-step map is close to a low-degree polynomial, and validation reduces
compact \gscn{} to its polynomial trend (\S\ref{sec:field}), at no cost: on Lorenz-63 with 400, 2{,}000 and 10{,}000
observations it forecasts \num{l63.400.ChaosSplat}, \num{l63.2000.ChaosSplat} and \num{l63.10000.ChaosSplat} Lyapunov
times, never significantly behind a baseline except NG-RC searched over its own, longer windows with 10{,}000 points
(\num{l63.10000.NGRCclosedloopselection}); on Lorenz-96 with $d=5$ to 40, \num{l96.5.ChaosSplat},
\num{l96.10.ChaosSplat}, \num{l96.20.ChaosSplat} and \num{l96.40.ChaosSplat}, exactly as NG-RC selected the same
way. Validation rejects the splats on Lorenz-63 and on Lorenz-96 with $d=5$, 10; from $d=20$ they exceed the
memory limit and are not tried. The window of past states is selected by validation forecast horizon
(\ssecref{sec:closedloop}); baselines in Table~\sref{tab:polymodels}; Wilcoxon signed-rank tests, Holm-corrected. Why
the polynomial forecasters differ, and PolyProp's 512-bit fit under float32 data: \ssecref{sec:smooth-full}.

%% ====================================================================
%% Fresh confirmation. Tables: experiments/overleaf/make_v4_tables.py from fresh_v4/results (initial conditions A, B = 9, 10 in the code).
%% ====================================================================

\section{Fresh pre-registered confirmation}\label{sec:fresh}

Full mode was frozen at a repository commit, with its claims, before its data existed, and tested on the ten systems
of \S\S\ref{sec:maps}--\ref{sec:heldout} at $n=10^4$ and $10^5$ on two fresh initial conditions, A and B, against the
S-map and the Farmer--Sidorowich predictor with the inputs of \S\ref{sec:fairness}. The units are series
3 and R\"ossler (smooth; R\"ossler at $10^4$ only) and eight kinked maps; a unit is one system at one training size and
one initial condition. Before every run the scripts check that the method files match the frozen commit. Claims use the
one-step MSE on the common test set $t\ge80$ with a 95\% moving-block bootstrap interval of the error ratio (blocks of
500 points, 2\,000 resamples). Runs used Kaggle's cloud machines with one linear-algebra thread (the third initial
condition below, the authors' workstation); the CPU type is recorded per job. One baselines job (R\"ossler, A) hit
Kaggle's 12-hour limit and was completed with the same frozen code, as the protocol prescribed. The record of the
freeze is the repository's commit history, pushed to GitHub as each step happened, not an independent registry. \gscn{} beat both on
\num{f910.unitwins} of \num{f910.units} units (\num{f910.resolved} resolved) and on \num{f910.syswins} of
\num{f910.sys} systems at a given size, with a median error \num{f910.ratio.medianinv} times lower
(Fig.~\ref{fig:fresh}, Table~\sref{tab:fresh910}), R\"ossler included (ratios 0.40 and 0.81
against the Farmer--Sidorowich predictor), where compact mode loses (\ssecref{sec:flows}). Units of one system share its
dynamics, so they are not independent: by system, every unit of all \num{sys.n} systems was a resolved win, and the
geometric-mean error was \num{sys.inv.min} (\num{sys.inv.min.name}) to $\numm{sys.inv.max}$ times lower,
\num{sys.inv.median} at the median. Over 5 and 20
closed-loop steps (reported, not claimed) \gscn{} was the more accurate on \num{f910.roll5.wins} and
\num{f910.roll20.wins} of \num{f910.roll.units} units. A single seed's colour-channel composite won
\num{f910.oneseed.wins} of \num{f910.oneseed.units}.

\ifjournal\else %% SUPP-BEGIN Fresh confirmation: table

\fi %% SUPP-END

\paragraph{A third initial condition} On C, declared before its data existed and run with the frozen code and scoring,
\gscn{} beat both baselines on \num{f11.unitwins} of \num{f11.units} units (\num{f11.resolved} resolved), with a median
error \num{f11.ratio.medianinv} times lower (Table~\sref{tab:fresh11}); a single seed's composite won
\num{f11.oneseed.wins} of \num{f11.oneseed.units}; over A, B and C, \gscn{} won \num{f911.unitwins} of \num{f911.units}
units.\footnote{C was added after the confirmation so that the claim does not rest on two trajectories per system; its
offsets were checked against every one used before, and it ran on the authors' workstation instead of Kaggle. These
counts are reported beside the pre-registered claims for A and B, not in their place.}

\paragraph{Components and robustness} Each component added in turn raised the wins (Table~\sref{tab:ladder910}), and the
six-member combination removed the colour-channel composites' \num{f910.vsfinal.big} largest errors. The referee alone won \num{f910.netalone.wins} of \num{f910.netalone.units} kinked units. The combination removed those
errors by factors of $\numm{f910.vsfinal.mingain}$ to $\numm{f910.vsfinal.maxgain}$, at a cost of at most a factor of
\num{f910.vsfinal.maxloss} on the other units.

On development data \gscn{} was within a factor of 1.04 of the exact map, or below it, on 15 of 16 units in each
unit's worst floating-point environment, and varied by at most a factor of 2.6 across three environments.\footnote{The
exception, the border-collision map at initial condition 1, at 116 times, is still $2\times10^5$ times below the best
baseline (Table~\sref{tab:v4-robust}).}

\ifjournal\else %% SUPP-BEGIN Development robustness

\fi %% SUPP-END

\begin{figure*}[tbp]
\centering
\includegraphics[width=\textwidth]{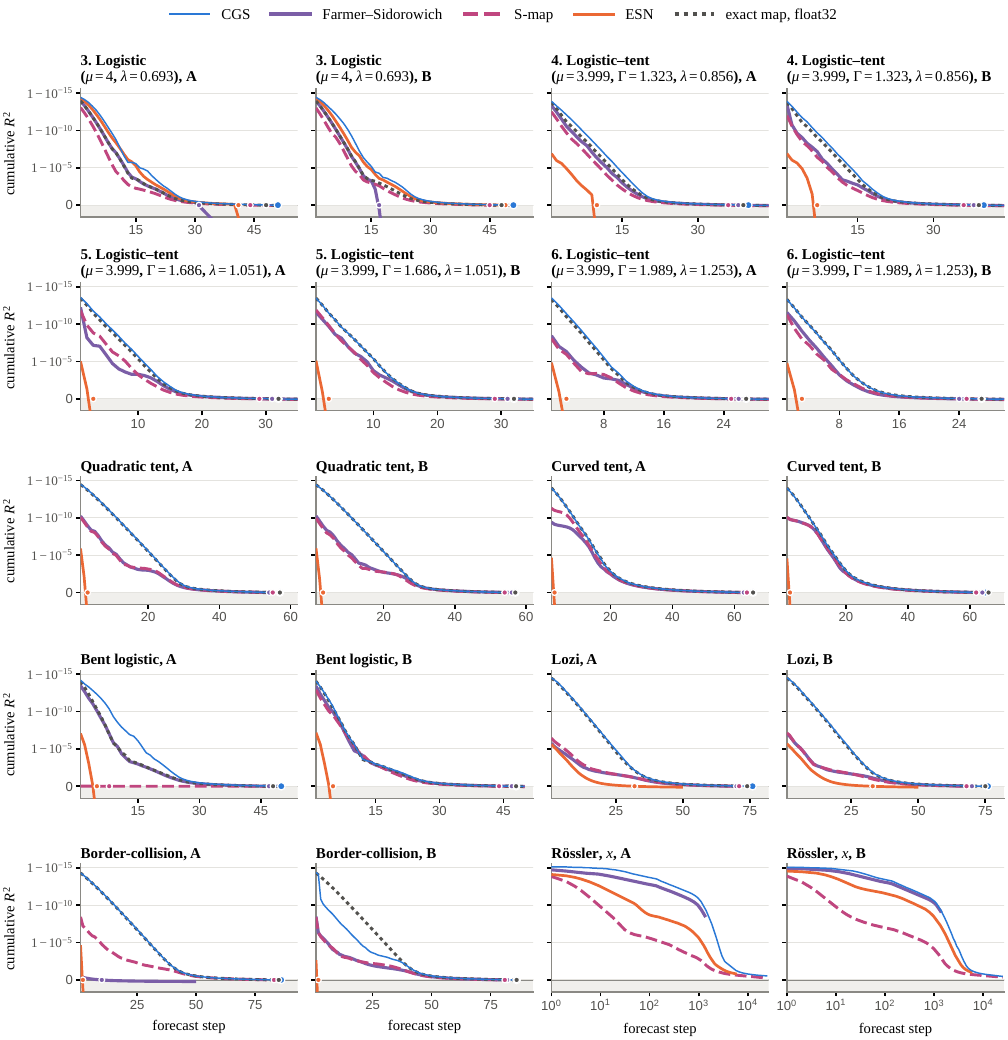}
\caption{Closed-loop forecasts on the fresh units ($n=10^4$, initial conditions A and B) until each model's cumulative
$R^2$ over 3\,000 test windows has fallen to 0 (dots; float32 feedback). The vertical axis shows $1-R^2_{\mathrm{cum}}$
on a log scale, labelled by $R^2_{\mathrm{cum}}$; below 0, shaded. The exact map, applied to the float32 observations,
is the reference on the maps. On R\"ossler (log horizontal axis) the forecasts ran to 25\,000 steps (the Farmer--Sidorowich
predictor's, which are slow, to about 1\,400, and the ESN's to 6\,000), and no model's cumulative $R^2$ reached 0 within them. Steps in Table~\sref{tab:fresh-rollouts}.}
\label{fig:fresh-rollouts}
\end{figure*}

\begin{figure*}[!htbp]
\centering
\includegraphics[width=\textwidth]{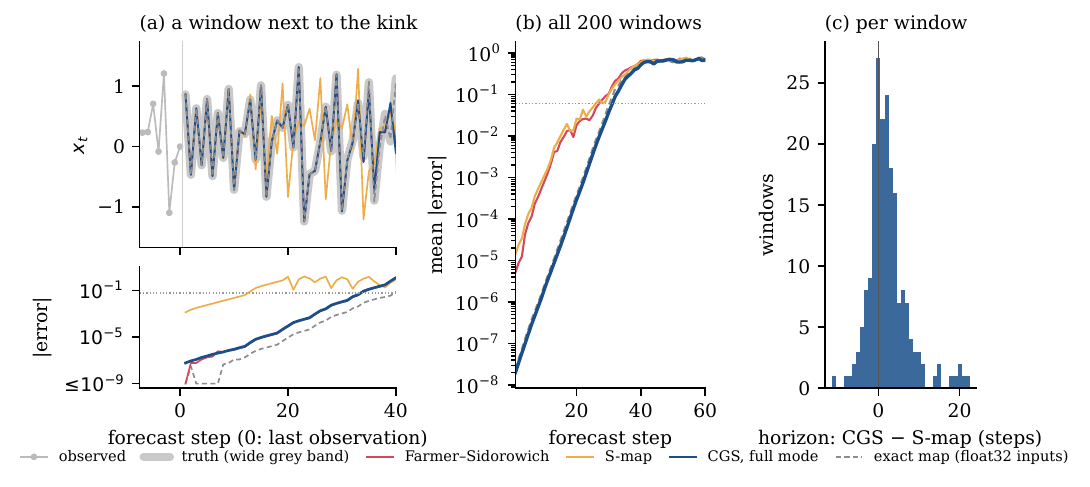}
\caption{Closed-loop forecasts on a fresh kinked unit (Lozi, initial condition A, $n=10^4$; post hoc, reported, not
claimed): full-mode \gscn{}, the S-map, the Farmer--Sidorowich predictor and the exact map on the float32 inputs,
fed back in float32 from \num{traj.windows} test windows. (a)~The window whose
first step the S-map misses most, next to the kink: forecasts over the truth (grey band) and, below, their errors;
\gscn{}'s starts at the float32 resolution and grows like the exact map's. (b)~Mean error at each step over all windows
(dotted: 10\% of the series' standard deviation). (c)~\gscn{}'s valid horizon minus the S-map's, per window: longer in
\num{traj.longer}, shorter in \num{traj.shorter}. Typically every model, the exact map included, holds for about \num{traj.v4.valid}
steps, the limit set by float32 feedback.}
\label{fig:lozi-trajectory}
\end{figure*}

\paragraph{Closed loop to the end} Run until the cumulative $R^2$ fell to 0 (post hoc), \gscn{}'s forecasts lasted
longer than the S-map's, the Farmer--Sidorowich predictor's and the ESN's on all \num{roll.full.units} map units, by
\num{roll.full.lead.min} to \num{roll.full.lead.max} steps, and as long as the exact map's (\num{roll.full.exact.min}
to \num{roll.full.exact.max} times its horizon; \ifjournal Fig.~\ref{fig:fresh-rollouts}\else Figs.~\ref{fig:fresh-rollouts} and~\ref{fig:lozi-trajectory}\fi). These horizons come from
3\,000 test windows, $n=10^4$; per-window errors were not stored, so these margins carry no interval. On R\"ossler
\gscn{}'s cumulative $R^2$ was still \num{roll.ross.v4.A} and \num{roll.ross.v4.B} after \num{roll.ross.cap} steps,
against \num{roll.ross.smap.A} and \num{roll.ross.smap.B} for the S-map, and it also led the ESN and the
Farmer--Sidorowich predictor over the shorter runs they allowed (\ssecref{sec:fresh-full}).

\ifjournal\else %% SUPP-BEGIN Full closed-loop rollouts on the fresh initial conditions

\fi %% SUPP-END

\paragraph{Cost} With one thread \gscn{} trained in a median of \num{f910.time.1e4.cs} minutes at $n=10^4$ and
\num{f910.time.1e5.cs} hours at $10^5$, against \num{f910.time.1e4.base} minutes and \num{f910.time.1e5.base} hours for
the baselines with all their input views, and predicted in \num{f910.pred.1e4.cs}\,$\mu$s per query against
\num{f910.pred.1e4.base}\,$\mu$s. The baselines on their own window alone took \num{f910.time.1e4.own} seconds and
\num{f910.time.1e5.own} minutes; scaling with $n$: Fig.~\ref{fig:scaling}.

\ifjournal\else %% SUPP-BEGIN Component ablation on the fresh initial conditions

\fi %% SUPP-END

%% Fairness of the comparison (experiments/paper_fairness)
\section{Fairness of the comparison}\label{sec:fairness}

The comparison guards against favourable inputs, a larger tuning budget, test-fitted development choices, hidden
information and selective reporting (Table~\sref{tab:criticisms}).\footnote{A corrected validation gap: in the NG-RC
and ESN code shared by the learning curves, the held-out tests and the rollouts, a few validation windows of models
trained below full size spanned a jump in time; the one-line fix and the reruns of every affected entry are described
in \ssecref{sec:fairness-full}. No claim changed.} Every unit is reported, losses and timed-out jobs included. The baselines get every input
view \gscn{} may use (their own window up to $m=16$, while \gscn{} is capped at 8, SVD embeddings and lagged views),
chosen by the same rule, with grids extended past any edge; \gscn{} has no per-system settings and fixed seeds (0, 1,
2). Against the best of all baseline variants on test, a choice that favours the baselines, \gscn{} still wins all 38
units, with a median error 38 times lower. Our S-map reproduces pyEDM, the reference implementation of
Sugihara's group, with the same settings (to 4--13 digits on 27 of 38 units; with its validated ridge it is the more
accurate on 9 and the less on 2), and pyEDM's default, with every training point in each fit, was less accurate than
ours on 13 of the 20 units at $n=10^4$ and more on 3. On 7 units the view rule chose a baseline variant less accurate on
test than the baseline on its own window. The best variants range over both baselines with and without views and
pyEDM's default; every interval is resolved where per-point errors exist; the narrowest margins are on series 3
(4--8\%).

\paragraph{Further baselines} After the confirmation, \num{extra.methods} further methods, each choosing its window and
settings on validation, were run on the same \num{extra.units} fresh units at $n=10^4$: \gscn{} had the lowest one-step
error on all \num{extra.cs.best}. The further methods are NG-RC, the ESN, SETAR, gradient-boosted trees, a random forest, MARS, a ReLU
network, Gaussian-process regression, the pre-trained chaos forecaster Panda~\cite{lai2026panda} and
TimeGS~\cite{wang2026forecasting}. Tests for hidden information (no data file opened during fitting, predictions
unchanged when other rows of a batch are replaced, time-shuffled series unpredictable, no test stretch in training) are
in \ssecref{sec:hidden}. PolyProp~\cite{polyprop2025}, with its authors' 512-bit code, was less accurate on
\num{ppf.wins} of \num{ppf.units} (on the kinked maps $\numm{ppf.kink.min}$ to $\numm{ppf.kink.max}$ times), and the
local model networks LOLIMOT and LOLIMOT-Q~\cite{nelles1996basis} on all \num{lol.wins} units of A, B and C
(\ssecref{sec:fairness-more}). Disclosures: the hinge parsimony margin (0.8) and several development decisions
used the test segments of development initial conditions; only A and B test \gscn{}. A predecessor without the
combination and the referee was confirmed on two earlier fresh initial conditions (\num{f78.unitwins} of
\num{f78.units} units); an earlier candidate's confirmation was stopped after one unit. Full mode's results on the
paper's own series and the closed-loop rollouts were run after the confirmation with its frozen code on development
data, on the machines available (CPU recorded per job), and are reported, not tested. Every variant tried is listed in
the repository.

\ifjournal\else %% SUPP-BEGIN Tests for hidden information: table

\fi %% SUPP-END

\ifjournal\else %% SUPP-BEGIN Anticipated criticisms: table

\fi %% SUPP-END

\section{Discussion}\label{sec:discussion}

\paragraph{Relation to RBF networks and partition-of-unity methods} Gaussian units with local polynomial models
blended by normalised weights are established: radial-basis-function networks~\cite{broomhead1988multivariable},
moving least squares~\cite{lancaster1981surfaces}, Takagi--Sugeno models~\cite{takagi1985fuzzy} and local model
networks~\cite{nelles1996basis}. Adapting their number and placement to the residual follows 3D Gaussian splatting.
What is specific to \gscn{} is the hinges, the location of kinks in the training windows' neighbour graph and the
referee that uses them; on the fresh units each component added in turn raised the wins
(Table~\sref{tab:ladder910}).

\paragraph{What the kinks change} A local fit near a kink is exact on either side but errs within one neighbourhood of
it: with $k$ neighbours among $n$ points in $D$ dimensions, and a data density bounded above and below near the kink, the neighbourhood has radius $h\propto(k/n)^{1/D}$, the
error there is of order $h$ on a band of width of order $h$, so its mean squared error falls only as $h^3\propto n^{-3/D}$ and concentrates where the dynamics
bends: on the \num{kb.units} kinked fresh units, the 1\% of test points nearest the kink carried a median
\num{kb.base.median}\% of the better baseline's squared error (more than half on \num{kb.base.majority}), and at most
\num{kb.cs.max}\% of full mode's (Fig.~\sref{fig:kink-errors}). Chaos then amplifies what remains: a forecast stays
within a tolerance $\delta$ for roughly $\ln(\delta/\varepsilon_0)/\lambda$ steps from an initial error $\varepsilon_0$,
an approximation that depends on the data's precision and on the criterion, so a one-step MSE $10^3$ times smaller
buys about $\ln(10^3)/(2\lambda)$ steps. Compact mode's cumulative $R^2$ reached zero after a number of steps times
$\lambda$ of 34 to 37 on series 2--6 and H\'enon (Table~\sref{tab:longr2}), as this predicts, and full mode's post hoc
lead of \num{roll.full.lead.min} to \num{roll.full.lead.max} steps (\S\ref{sec:fresh}) is of the same size.

\paragraph{Two modes, two trade-offs} The accuracy claims are full mode's and the memory claims compact mode's.
Compact mode stores a fixed model; full mode keeps the training windows for its referee, trains for
\num{f910.time.1e5.cs} hours at $n=10^5$ and predicts in \num{f910.pred.1e4.cs}\,$\mu$s per query, against
\num{f910.pred.1e4.base}\,$\mu$s for the local methods (\S\ref{sec:fresh}): no one configuration is at once the
most accurate, the smallest and the fastest. The results concern deterministic, low-dimensional series observed with
little noise; under observation noise the advantage shrinks (\S\ref{sec:noise}), and on the measured Santa Fe laser
compact mode and the S-map are level.

\paragraph{Limitations}\label{sec:limits} The splats do not overcome the curse of dimensionality: delay windows here have
at most 16 values, and on Lorenz-96 the splats exceed the memory budget from $d=20$. Where candidates are close on validation, a single
training depends on the numerical library (\ssecref{sec:machine}); full mode's combination limits this to 30\% across
CPU types on the fresh kinked maps.

%% ====================================================================

  % Table A.1, ...

%% ====================================================================
\ifjournal
\section*{CRediT authorship contribution statement}
\textbf{Daniel Ortiz Valete:} Conceptualization, Software. \textbf{Gonzalo Travieso:} Supervision, Writing -- review
\& editing. \textbf{Odemir Martinez Bruno:} Supervision, Writing -- review \& editing.

\section*{Declaration of competing interest}
The authors declare that they have no known competing financial interests or personal relationships that
could have appeared to influence the work reported in this paper.
\fi

%\section*{Data availability}
%All code, benchmark generators, result files and the scripts that produce every table and figure are
%openly available at \url{https://github.com/valeteda/CGS} (branch \nolinkurl{ld-cgs}). The supplementary material (S1, S2, \dots) is the ancillary file of the arXiv
%version and is also in the repository's \nolinkurl{manuscript} folder. Every table, and most numbers in the text, are written from the result files by the scripts in
%\nolinkurl{paper/} and \nolinkurl{experiments/overleaf/}, which also rebuild the manuscript (the other numbers are taken from those tables and files); the confirmation,
%hidden-information-test and further-baseline results sit beside the scripts that produced them. The pre-registration protocols,
%the dated record of the freeze and of the extension, and the record of every variant tried are in the same
%repository.

\section*{Acknowledgements}
D. O. Valete acknowledges funding from the Higher Education Personnel Improvement Coordination (CAPES) (\#88887.365931.2026-00), G. Travieso from the São Paulo Research
Foundation (FAPESP) (\#2026/15159-5), O. M. Bruno from FAPESP (\#2026/15159-5 and \#2025/27044-5) and Brazilian National Council for Scientific and Technological Development
(CNPq) (\#305610/2022-8).

\FloatBarrier
\clearpage
\appendix\setcounter{table}{0}\renewcommand{\dblfloatpagefraction}{0.99}\renewcommand{\dbltopfraction}{0.92}\makeatletter\setlength{\@dblfptop}{0pt}\setlength{\@dblfpsep}{12pt}\makeatother
\makeatletter\@addtoreset{figure}{section}\@addtoreset{table}{section}\makeatother\setcounter{figure}{0}  % Fig. B.1, Table B.1, ... in each appendix
\FloatBarrier

\begin{thebibliography}{10}
\expandafter\ifx\csname url\endcsname\relax
  \def\url#1{\texttt{#1}}\fi
\expandafter\ifx\csname urlprefix\endcsname\relax\def\urlprefix{URL }\fi
\expandafter\ifx\csname href\endcsname\relax
  \def\href#1#2{#2} \def\path#1{#1}\fi

\bibitem{kerbl20233d}
B.~Kerbl, G.~Kopanas, T.~Leimk{\"u}hler, G.~Drettakis, {3D} {G}aussian
  splatting for real-time radiance field rendering, ACM Transactions on
  Graphics 42~(4) (2023) 1--14.

\bibitem{ott2002chaos}
E.~Ott, Chaos in Dynamical Systems, 2nd Edition, Cambridge University Press,
  2002.

\bibitem{lorenz1969predictability}
E.~N. Lorenz, Predictability of a flow which possesses many scales of motion,
  Tellus 21 (1969) 289--307.

\bibitem{dibernardo2008piecewise}
M.~di~Bernardo, C.~J. Budd, A.~R. Champneys, P.~Kowalczyk, Piecewise-smooth
  Dynamical Systems: Theory and Applications, Vol. 163 of Applied Mathematical
  Sciences, Springer, London, 2008.
\newblock \href {https://doi.org/10.1007/978-1-84628-708-4}
  {\path{doi:10.1007/978-1-84628-708-4}}.

\bibitem{polyprop2025}
C.~Sch{\"o}tz, N.~Boers, Machine-precision prediction of low-dimensional
  chaotic systems from noise-free data, Nonlinear Dynamics 114 (2026) 1000.
\newblock \href {https://doi.org/10.1007/s11071-026-12915-9}
  {\path{doi:10.1007/s11071-026-12915-9}}.

\bibitem{hochreiter1997long}
S.~Hochreiter, J.~Schmidhuber, Long short-term memory, Neural Computation 9
  (1997) 1735--1780.

\bibitem{lai2018modeling}
G.~Lai, W.-C. Chang, Y.~Yang, H.~Liu, Modeling long- and short-term temporal
  patterns with deep neural networks, in: Proceedings of the 41st International
  ACM SIGIR Conference on Research and Development in Information Retrieval,
  2018, pp. 95--104.

\bibitem{vlachas2018data}
P.~R. Vlachas, W.~Byeon, Z.~Y. Wan, T.~P. Sapsis, P.~Koumoutsakos, Data-driven
  forecasting of high-dimensional chaotic systems with long short-term memory
  networks, Proceedings of the Royal Society A 474 (2018) 20170844.

\bibitem{vlachas2020backpropagation}
P.~R. Vlachas, J.~Pathak, B.~R. Hunt, T.~P. Sapsis, M.~Girvan, E.~Ott,
  P.~Koumoutsakos, Backpropagation algorithms and reservoir computing in
  recurrent neural networks for the forecasting of complex spatiotemporal
  dynamics, Neural Networks 126 (2020) 191--217.

\bibitem{valle2025forecasting}
J.~Valle, O.~M. Bruno, Forecasting chaotic time series: Comparative performance
  of {LSTM}-based and {T}ransformer-based neural network, Chaos, Solitons \&
  Fractals 192 (2025) 116034.

\bibitem{jaeger2004harnessing}
H.~Jaeger, H.~Haas, Harnessing nonlinearity: Predicting chaotic systems and
  saving energy in wireless communication, Science 304 (2004) 78--80.

\bibitem{pathak2017using}
J.~Pathak, Z.~Lu, B.~R. Hunt, M.~Girvan, E.~Ott, Using machine learning to
  replicate chaotic attractors and calculate {L}yapunov exponents from data,
  Chaos 27 (2017) 121102.

\bibitem{pathak2018model}
J.~Pathak, B.~Hunt, M.~Girvan, Z.~Lu, E.~Ott, Model-free prediction of large
  spatiotemporally chaotic systems from data: A reservoir computing approach,
  Physical Review Letters 120 (2018) 024102.

\bibitem{gauthier2021next}
D.~J. Gauthier, E.~Bollt, A.~Griffith, W.~A.~S. Barbosa, Next generation
  reservoir computing, Nature Communications 12 (2021) 5564.

\bibitem{brunton2016discovering}
S.~L. Brunton, J.~L. Proctor, J.~N. Kutz, Discovering governing equations from
  data by sparse identification of nonlinear dynamical systems, Proceedings of
  the National Academy of Sciences 113~(15) (2016) 3932--3937.
\newblock \href {https://doi.org/10.1073/pnas.1517384113}
  {\path{doi:10.1073/pnas.1517384113}}.

\bibitem{ansari2024chronos}
A.~F. Ansari, L.~Stella, C.~Turkmen, X.~Zhang, P.~Mercado, H.~Shen, O.~Shchur,
  S.~S. Rangapuram, S.~Pineda~Arango, S.~Kapoor, et~al., Chronos: Learning the
  language of time series, Transactions on Machine Learning Research (2024).

\bibitem{zhang2025zeroshot}
Y.~Zhang, W.~Gilpin, Zero-shot forecasting of chaotic systems, in:
  International Conference on Learning Representations (ICLR), 2025,
  arXiv:2409.15771.

\bibitem{takens1981detecting}
F.~Takens, Detecting strange attractors in turbulence, in: Dynamical Systems
  and Turbulence, Vol. 898 of Lecture Notes in Mathematics, Springer, 1981, pp.
  366--381.

\bibitem{letellier1998nonequivalence}
C.~Letellier, J.~Maquet, L.~Le~Sceller, G.~Gouesbet, L.~A. Aguirre, On the
  non-equivalence of observables in phase-space reconstructions from recorded
  time series, Journal of Physics A: Mathematical and General 31~(39) (1998)
  7913--7927.

\bibitem{farmer1987predicting}
J.~D. Farmer, J.~J. Sidorowich, Predicting chaotic time series, Physical Review
  Letters 59 (1987) 845--848.

\bibitem{casdagli1989nonlinear}
M.~Casdagli, Nonlinear prediction of chaotic time series, Physica D 35 (1989)
  335--356.

\bibitem{sugihara1994nonlinear}
G.~Sugihara, Nonlinear forecasting for the classification of natural time
  series, Philosophical Transactions of the Royal Society A 348 (1994)
  477--495.

\bibitem{westover1990footprint}
L.~Westover, Footprint evaluation for volume rendering, ACM SIGGRAPH Computer
  Graphics 24~(4) (1990) 367--376.
\newblock \href {https://doi.org/10.1145/97880.97919}
  {\path{doi:10.1145/97880.97919}}.

\bibitem{wang2026forecasting}
Y.~Wang, Y.~Hu, P.~Liu, N.~Li, T.~Dai, S.-T. Xia, Forecasting as rendering: A
  {2D} {G}aussian splatting framework for time series forecasting, in:
  Proceedings of the 32nd ACM SIGKDD Conference on Knowledge Discovery and Data
  Mining V.2, 2026, pp. 5160--5170, arXiv:2603.02220.

\bibitem{broomhead1988multivariable}
D.~S. Broomhead, D.~Lowe, Multivariable functional interpolation and adaptive
  networks, Complex Systems 2~(3) (1988) 321--355.

\bibitem{platt1991resource}
J.~C. Platt, A resource-allocating network for function interpolation, Neural
  Computation 3~(2) (1991) 213--225.
\newblock \href {https://doi.org/10.1162/neco.1991.3.2.213}
  {\path{doi:10.1162/neco.1991.3.2.213}}.

\bibitem{stokbro1990exploiting}
K.~Stokbro, D.~K. Umberger, J.~A. Hertz, Exploiting neurons with localized
  receptive fields to learn chaos, Complex Systems 4~(6) (1990) 603--622.

\bibitem{takagi1985fuzzy}
T.~Takagi, M.~Sugeno, Fuzzy identification of systems and its applications to
  modeling and control, IEEE Transactions on Systems, Man, and Cybernetics
  SMC-15~(1) (1985) 116--132.
\newblock \href {https://doi.org/10.1109/TSMC.1985.6313399}
  {\path{doi:10.1109/TSMC.1985.6313399}}.

\bibitem{nelles1996basis}
O.~Nelles, R.~Isermann, Basis function networks for interpolation of local
  linear models, in: Proceedings of the 35th IEEE Conference on Decision and
  Control, Vol.~1, 1996, pp. 470--475.
\newblock \href {https://doi.org/10.1109/CDC.1996.574356}
  {\path{doi:10.1109/CDC.1996.574356}}.

\bibitem{vijayakumar2005incremental}
S.~Vijayakumar, A.~D'Souza, S.~Schaal, Incremental online learning in high
  dimensions, Neural Computation 17~(12) (2005) 2602--2634.
\newblock \href {https://doi.org/10.1162/089976605774320557}
  {\path{doi:10.1162/089976605774320557}}.

\bibitem{daniels2026splat}
M.~Daniels, P.~Rigollet, Splat regression models, in: Proc. Int. Conf. Learning
  Representations (ICLR), 2026, arXiv:2511.14042.

\bibitem{breiman1993hinging}
L.~Breiman, Hinging hyperplanes for regression, classification, and function
  approximation, IEEE Transactions on Information Theory 39~(3) (1993)
  999--1013.
\newblock \href {https://doi.org/10.1109/18.256506}
  {\path{doi:10.1109/18.256506}}.

\bibitem{friedman1991multivariate}
J.~H. Friedman, Multivariate adaptive regression splines, The Annals of
  Statistics 19~(1) (1991) 1--67.
\newblock \href {https://doi.org/10.1214/aos/1176347963}
  {\path{doi:10.1214/aos/1176347963}}.

\bibitem{paoletti2007identification}
S.~Paoletti, A.~L. Juloski, G.~Ferrari-Trecate, R.~Vidal, Identification of
  hybrid systems: a tutorial, European Journal of Control 13~(2--3) (2007)
  242--260.
\newblock \href {https://doi.org/10.3166/ejc.13.242-260}
  {\path{doi:10.3166/ejc.13.242-260}}.

\bibitem{kaheman2020sindy}
K.~Kaheman, J.~N. Kutz, S.~L. Brunton, {SINDy-PI}: a robust algorithm for
  parallel implicit sparse identification of nonlinear dynamics, Proceedings of
  the Royal Society A 476~(2242) (2020) 20200279.
\newblock \href {https://doi.org/10.1098/rspa.2020.0279}
  {\path{doi:10.1098/rspa.2020.0279}}.

\bibitem{li2026image}
W.~Li, X.-Y. Wang, Z.~Zhu, X.~Zhang, L.~Zhang, Image encryption via
  data-identified discrete chaotic maps, arXiv:2605.21118 (2026).

\bibitem{henon1976two}
M.~H{\'e}non, A two-dimensional mapping with a strange attractor,
  Communications in Mathematical Physics 50 (1976) 69--77.

\bibitem{lorenz1963deterministic}
E.~N. Lorenz, Deterministic nonperiodic flow, Journal of the Atmospheric
  Sciences 20 (1963) 130--141.

\bibitem{rossler1976equation}
O.~E. R{\"o}ssler, An equation for continuous chaos, Physics Letters A 57
  (1976) 397--398.

\bibitem{weigend1994time}
A.~S. Weigend, N.~A. Gershenfeld (Eds.), Time Series Prediction: Forecasting
  the Future and Understanding the Past, Addison-Wesley, Reading, MA, 1994.

\bibitem{gilpin2021chaos}
W.~Gilpin, Chaos as an interpretable benchmark for forecasting and data-driven
  modelling, in: Proceedings of the Neural Information Processing Systems Track
  on Datasets and Benchmarks, 2021.

\bibitem{lorenz1996predictability}
E.~N. Lorenz, Predictability: A problem partly solved, in: Proceedings of the
  Seminar on Predictability, ECMWF, Reading, 1996, pp. 1--18.

\bibitem{marwan2009complex}
N.~Marwan, J.~F. Donges, Y.~Zou, R.~V. Donner, J.~Kurths, Complex network
  approach for recurrence analysis of time series, Physics Letters A 373 (2009)
  4246--4254.

\bibitem{donner2010recurrence}
R.~V. Donner, Y.~Zou, J.~F. Donges, N.~Marwan, J.~Kurths, Recurrence
  networks---a novel paradigm for nonlinear time series analysis, New Journal
  of Physics 12 (2010) 033025.

\bibitem{small2013complex}
M.~Small, Complex networks from time series: Capturing dynamics, in:
  Proceedings of the IEEE International Symposium on Circuits and Systems,
  2013, pp. 2509--2512.

\bibitem{bernstein1914}
S.~Bernstein, Sur la meilleure approximation de $|x|$ par des polyn{\^o}mes de
  degr{\'e}s donn{\'e}s, Acta Mathematica 37 (1914) 1--57.

\bibitem{lozi1978attracteur}
R.~Lozi, Un attracteur {\'e}trange (?) du type attracteur de {H}{\'e}non,
  Journal de Physique Colloques 39~(C5) (1978) 9--10.
\newblock \href {https://doi.org/10.1051/jphyscol:1978505}
  {\path{doi:10.1051/jphyscol:1978505}}.

\bibitem{banerjee1998robust}
S.~Banerjee, J.~A. Yorke, C.~Grebogi, Robust chaos, Physical Review Letters
  80~(14) (1998) 3049--3052.
\newblock \href {https://doi.org/10.1103/PhysRevLett.80.3049}
  {\path{doi:10.1103/PhysRevLett.80.3049}}.

\bibitem{lai2026panda}
J.~Lai, A.~Bao, W.~Gilpin, Panda: A pretrained forecast model for chaotic
  dynamics, in: International Conference on Learning Representations (ICLR),
  2026.

\bibitem{lancaster1981surfaces}
P.~Lancaster, K.~Salkauskas, Surfaces generated by moving least squares
  methods, Mathematics of Computation 37~(155) (1981) 141--158.
\newblock \href {https://doi.org/10.1090/S0025-5718-1981-0616367-1}
  {\path{doi:10.1090/S0025-5718-1981-0616367-1}}.

\end{thebibliography}
\end{document}